\documentclass[]{aa} 
\bibpunct{(}{)}{;}{a}{}{,}

\usepackage{rotating}
\usepackage{subfig}
\usepackage{graphicx}
\usepackage{txfonts}
\begin{document}
%
   %\title{Significant oblateness of a smooth Galactic stellar halo }
   \title{A skewer survey of the Galactic halo from deep CFHT and INT images}
   
   %\subtitle{(dull, temporary title)}

   \author{B. Pila-D\'iez \inst{1}
          %\inst{\ref{inst1}}
	  \and
	  J.T.A. de Jong \inst{1}
	  \and
	  K. Kuijken \inst{1} 
          \and
	  R.F.J. van der Burg \inst{1,2}
          \and
	  H. Hoekstra \inst{1}}

   \institute{Leiden Observatory, Leiden University,
              Oort Building, Niels Bohrweg 2, NL-2333 CA Leiden\\
              \email{piladiez@strw.leidenuniv.nl}  \email{kuijken@strw.leidenuniv.nl}
	      \email{jelte@strw.leidenuniv.nl} \email{hoekstra@strw.leidenuniv.nl}
              \email{remco.van-der-burg@cea.fr}
              \label{inst1}
              \and
              Laboratoire AIM, IRFU/Service d’Astrophysique - CEA/DSM - CNRS, 
              Université Paris Diderot, Bat. 709, CEA-Saclay, 91191 Gif-sur-Yvette Cedex, France
            %  J.H. Oort Building, Niels Bohrweg 2, NL-2333 CA Leiden\\
         %     \email{jelte@strw.leidenuniv.nl}%\label{inst2}
              \label{inst2}
 	 %\and
            % Leiden Observatory, Leiden University,
            %  J.H. Oort Building, Niels Bohrweg 2, NL-2333 CA Leiden\\
         %     \email{kuijken@strw.leidenuniv.nl}%\label{inst2}
	 %\and
            % Leiden Observatory, Leiden University,
            %  J.H. Oort Building, Niels Bohrweg 2, NL-2333 CA Leiden\\
         %     \email{hoekstra@strw.leidenuniv.nl}%\label{inst2}
             %\thanks{The university of heaven temporarily does not
             %        accept e-mails}
             }

   \date{Received Month 00, 2014; accepted Month 00, Year}

% \abstract{}{}{}{}{} 
% 5 {} token are mandatory
% For traditional abstract, use just one pair of brackets (comment the rest).
 
  \abstract
   {We study the density profile and shape of the Galactic halo using deep multicolour 
   images from the MENeaCS and CCCP projects, over 33 fields selected to avoid overlap 
   with the Galactic plane. 
   Using multicolour selection and PSF homogenization techniques we obtain catalogues 
   of F stars (near-main sequence turnoff stars) out to Galactocentric distances up to 
   60kpc. Grouping nearby lines of sight, we construct the stellar density profiles 
   through the halo in eight different directions by means of photometric parallaxes. 
   Smooth halo models are then fitted to these profiles.
   We find clear evidence for a steepening of the density profile power law index around 
   $R=20$ kpc, from $-2.50\pm0.04$ to $-4.85\pm0.04$, and for a flattening of the halo 
   towards the poles with best-fit axis ratio $0.63\pm0.02$. Furthermore, we cannot rule 
   out a mild triaxiality ($w\geq0.8$). We recover the signatures of 
   well-known substructure and streams that intersect our lines of sight. These results 
   are consistent with those derived from wider but shallower surveys, and augur well for 
   upcoming, wide-field surveys of comparable depth to our pencil beam surveys. 
   }  
  % {Leave empty.}
  % {Leave empty.}
  % {Leave empty.}

   \keywords{Galaxy: halo, Galaxy: structure
               }
   % A maximum of 6 key words should be listed after the abstract. These must be selected from a list that is published each year in the first issue in January and is also available in Appendix A or on the A&A web site.

   \maketitle
%
%%%%%%%%%%%%%%%%%%%%%%%%%%%%%%%%%%%%%%%%%%%%%%%%%%%%%%%%%%%%%%%%%%%%%%%%%%%%%%%%%%%%%%%%%%%%
%  _________________________________________________________________________________________
% 
%  1. INTRODUCTION
%  _________________________________________________________________________________________
%
%%%%%%%%%%%%%%%%%%%%%%%%%%%%%%%%%%%%%%%%%%%%%%%%%%%%%%%%%%%%%%%%%%%%%%%%%%%%%%%%%%%%%%%%%%%%

\section{Introduction}\label{intro}

The stellar halo of the Milky Way only contains a tiny fraction of its stars, yet it provides important clues about the formation of the Galaxy and galaxy formation in general. Within the paradigm of hierarchical structure formation, galaxies evolve over time, growing by means of mergers and accretion of smaller systems. While in the central parts of galaxies the signatures of such events are rapidly dissipated, the long dynamical timescales allow accretion-induced substructures to linger for Gigayears in their outermost regions. Thus, the stellar structure of the outer halos of galaxies such as the Milky Way can help constrain not only the formation history of individual galaxies, but also cosmological models of structure formation.

Owing to the intrinsic faintness of stellar halos, the Milky Way is our best bet for a detailed study of such structures. However, even studying the Galactic stellar halo is fraught with difficulties; very sensitive data are required to probe stars at these large distances (out to ~100 kpc), and spread over sufficiently large areas to constrain the overall structure as well as localized substructures. In recent decades the advent of CCD-based all-sky surveys such as the Sloan Digital Sky Survey \citep[SDSS][]{york00,ahn14} in the optical and the 2 Micron All Sky Survey \citep[2MASS][]{skrutskie06} in the infrared have unlocked unprecedented views of the outer regions of the Galaxy. This has led to the discovery of many previously unknown substructures \citep[e.g.][]{newb02,belok06,grillmair06anticenter,belok07orphan,juric08,bell08} and to improved knowledge of the overall structure in these outskirts \citep[e.g.][]{chen01,juric08,dejong10,sesar10densProfile,sesar11,faccioli14}. Nevertheless, most of these recent analyses are still limited to either the inner parts of the stellar halo ($R_{GC} \leq 30$ kpc) or to particular, sparse stellar tracers (e.g. K-giants or RR Lyrae).

In this paper we use deep photometry obtained with the Canada-France-Hawaii Telescope (CFHT) MegaCam 
and the Wide Field Camera (WFC) at the Isaac Newton Telescope (INT), scattered over a large range of 
Galactic latitudes and longitudes to probe main sequence turn-off (MSTO) stars out to distances of 
60 kpc. Combining our data into eight independent lines of sight through the Galactic halo, we are able 
to constrain the overall structure of the outer halo, and to probe the substructure in these 
outermost regions. In section 2 we describe the data set used for this analysis and the construction 
of our deep star catalogues. Section 3 presents the derived stellar density profiles and smooth 
Galactic model fits. We discuss our results in section 4 and present our conclusions in section 5.

%%%%%%%%%%%%%%%%%%%%%%%%%%%%%%%%%%%%%%%%%%%%%%%%%%%%%%%%%%%%%%%%%%%%%%%%%%%%%%%%%%%%%%%%%%%%
%  _________________________________________________________________________________________
% 
%  2. OBSERVATIONS AND DATA PROCESSING
%  _________________________________________________________________________________________
%
%%%%%%%%%%%%%%%%%%%%%%%%%%%%%%%%%%%%%%%%%%%%%%%%%%%%%%%%%%%%%%%%%%%%%%%%%%%%%%%%%%%%%%%%%%%%

\section{Observations and data processing }\label{data}

   % ---------------------------------------------------------------------------------------
   % 2.1 DESCRIPTION OF DATA SET
   % ---------------------------------------------------------------------------------------

   \subsection{Survey and observations}\label{subsec:observations}
   
   We use $g$ and $r$ images from the MENeaCS and the CCCP surveys 
   \citep{sand12meneacs,hoekstra12cccp,bildfell12combined} together with several archival 
   cluster fields from the CFHT-MegaCam instrument. We combine these data with $U$ and $i$ 
   images from a follow-up campaign with the INT-WFC instrument (van der Burg et al., in prep.). 
   Whereas these surveys targeted a preselected sample of galaxy clusters, the pointings 
   constitute a "blind" survey of the Milky Way stellar halo since their distribution is 
   completely independent of any prior  knowledge of the halo's structure and substructure.

   % Figure: Equatorial map of pointings
   %*************************************
   \begin{figure*}
   \centering
   \includegraphics[width=\textwidth]{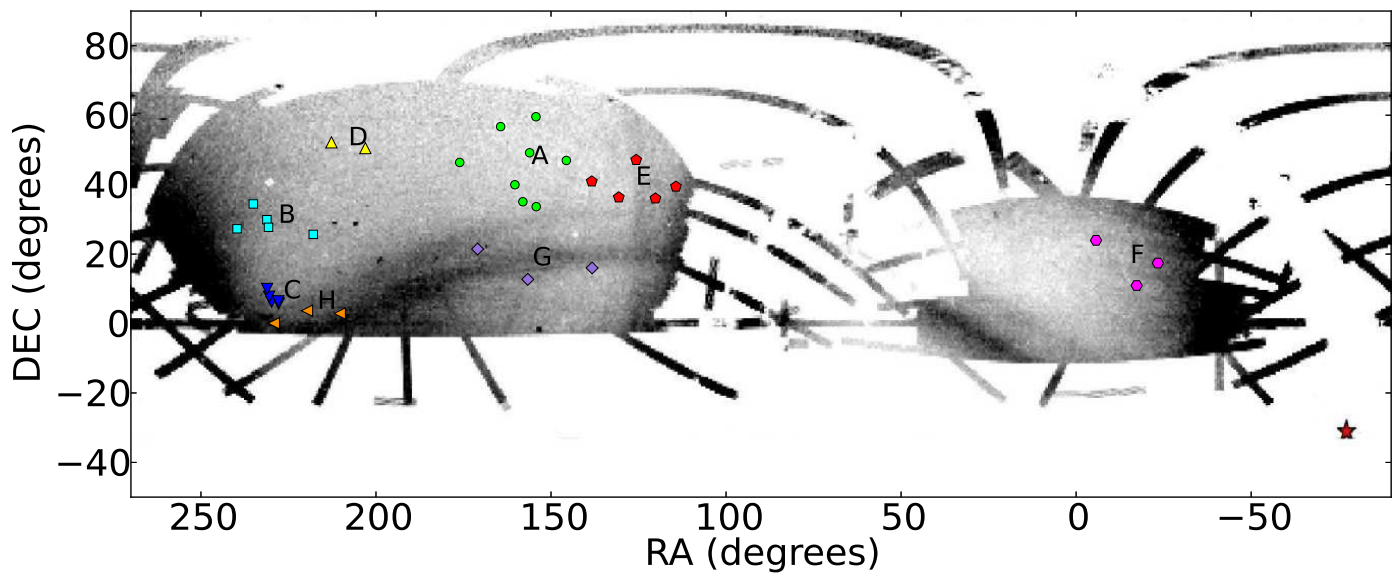}
      \caption{Equatorial map showing the position of all the fields used in this work. 
               The different colours and symbols indicate how the fields have been grouped 
               to calculate the different density profiles. 
               The background image is the SDSS-DR8 map from \citet{kopos12}, which shows the 
               footprint of the Sagittarius stream and the location of the Sagittarius dwarf 
               galaxy. When grouping the fields, we have also 
               taken into account the presence of this stream, the Triangulum-Andromeda overdensity, 
               and the anticentre substructures (ACS, EBS, and Monoceros), 
               in trying to combine their effect in certain profiles and avoid it in others.  
               }
         \label{fig:mapPointings}
   \end{figure*}

   Our pointings are distributed over the region of the sky visible to both the CFHT and the 
   INT (see Figure~\ref{fig:mapPointings}). To optimize the star-galaxy separation (see 
   section~\ref{subsec:PSF}) we restrict our 
   analysis to exposures with image quality of subarcsecond seeing, typically $<\approx0.9\ 
   \mathrm{arcsec}$ in the $r$ band. This limitation, combined with the varying fields of view and 
   observing conditions between the data sets, leads to pointing footprint sizes that range 
   between $0.24$ and $1.14\, \mathrm{deg}^2$. 
   %Our pointings are disseminated over the sky visible to both the CFHT and the INT.  Their  
   %initial g and r field-areas --covering between $0.6\, \mathrm{deg}^2$ and $3.7\, 
   %\mathrm{deg}^2$-- shrink to areas in the $(0.24,1.14)\, \mathrm{deg}^2$ range 
   %after combining them with the u and i footprints. This is due to two circumstances: first, 
   %the intrinsically smaller field-of-view of the INT and, second, the varying photometric 
   %quality of the i and u ditters observed throughout several months of uneven atmospheric 
   %conditions. To optimize 
   %the star-galaxy separation (see section~\ref{subsec:PSF}), we only use exposures with image 
   %quality of sub-arcsecond seeing, typically $<\approx0.9\ \mathrm{\textbf{arcsec}}$. 

   % ---------------------------------------------------------------------------------------
   % 2.2 PSF DISTORTION AND star-galaxy SEPARATION
   % ---------------------------------------------------------------------------------------

   \subsection{Image correction of the PSF distortion [and implications for 
      the star-galaxy separation]}\label{subsec:PSF}
 
   Previous research by our group has shown that the performance of standard star-galaxy 
   separation methods based on the size and ellipticity of the sources can be improved 
   by homogenizing the point-spread function (PSF) across an image prior to its photometric 
   analysis \citep{piladiez14}. In addition, such a correction also provides the benefit of 
   allowing us to perform fixed aperture photometry and colour measurements.

   In order to homogenize the PSF of our images, we use a code \citep{piladiez14} 
   that, as a first step, takes the shape of the bright stars in a given image and uses it 
   to map the varying PSF and, as a second step, convolves this map with a spatially variable 
   kernel designed to transform everywhere the original PSF into a gaussian PSF.

   % ---------------------------------------------------------------------------------------
   % 2.3 Catalogues
   % ---------------------------------------------------------------------------------------
   
   \subsection{Catalogues}\label{subsec:catalogues}
   
   From the PSF-homogenized exposures we create photometric catalogues using Source Extractor
   \citep{bertinSExtractor}. 
   For the $g$ and the $r$ data, we stack the different exposures in each band to create a single 
   calibrated image, and we extract the band catalogues from them. We perform a star-galaxy 
   separation based on the brightness, size and ellipticity of the sources and we match the 
   surviving sources in the two catalogues to produce a $gr$-catalogue of stars for each field 
   of view (see \citet{piladiez14}). The limiting magnitudes of these $gr$ star catalogues 
   reach $m_{AB}\sim25.0$ at the $5.0\sigma$ level in the $r$ band.
   
   For the $U$ and the $i$ fields of view, we produce several photometric catalogues, one for each 
   individual exposure. We correct the magnitudes 
   in the $i$ catalogues for the dependency of the illumination on pixel position. For each 
   pointing and band, the exposure catalogues are calibrated to a common zero point and 
   combined to produce a single-band catalogue. In these single-band catalogues, the 
   resulting magnitude for each source is calculated as the median of the contributions of 
   all the individual exposures. At this point the $U$ and the $i$ magnitudes are converted from 
   the INT to the CFHT photometric system using the following equations, which we derive by 
   calibrating our mixed INT-CFHT colours to the colour stellar loci of the CFHT Legacy Survey 
   (\citet{erben09}, \citet{hildebrandt09}):
   \begin{eqnarray}
      i_{MegaCam} &=& i_{INT} - 0.12*(r_{Mega} - i_{INT})  \\
      u_{MegaCam} &=& u_{INT} - 0.15*(u_{INT} - g_{Mega})  \,.
   \end{eqnarray} 

\begin{figure}
   \centering
   \includegraphics[width=\columnwidth]{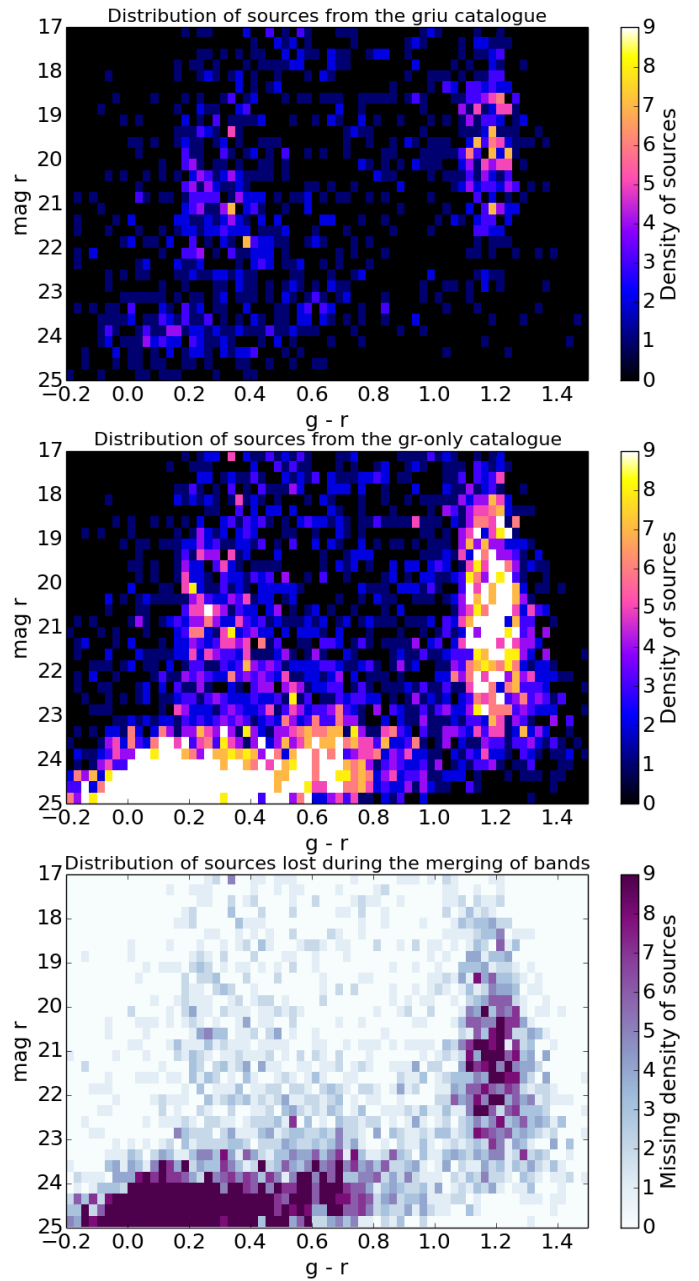}
      \caption{Hess diagrams showing the number of sources per colour-magnitude bin in the $ugri$
               catalogue (top), in the $gr$ catalogue (centre) and the difference between both 
               (bottom) for field A1033. Most of the sources lost when combining the catalogues 
               correspond to faint magnitudes, because the $i$ and the $U$ observations are shallower.
               The effect is the removal of most of the faint galaxies (located in the 
               $-0.2<g-r<0.7$ and $r>23$ region in the central panel), most of the faintest disk 
               M dwarves ($1.1<g-r<1.3$) and a number of faint objects (in the $i$ or 
               the $U$ bands) scattered throughout the $(g-r,r)$ diagram.
               }
         \label{fig:ugriCMD}
   \end{figure}

   Finally we position-match the sources from the $U$-, the $i$- and the $gr$-catalogues to 
   create a final catalogue of stellar %small round 
   sources for each field of view. These final $ugri$-catalogues are shallower than the 
   $gr$-catalogues because of the lesser depth of the $i$ and the $U$ observations (
   see Table~\ref{table:groups_of_view}). 
   %the limiting magnitudes now reach $\sim24.0$ mag at the $5.0\sigma$ level in the r band).  
   %The combination of the shallower i- and u-catalogues, however, improves 
   %the star-galaxy separation by removing galaxies too faint in the blue colours to survive 
   %the limiting magnitude of the CFHT survey but somewhat brighter in the red. 
   Figure~\ref{fig:ugriCMD} shows the colour-magnitude diagrams (CMDs) for the final $ugri$ 
   and $gr$ catalogues (top and centre, respectively), and the difference between them (bottom). 
   The bottom panel highlights that, in the colour regime of the halo ($0.2<g-r<0.3$), the 
   combination of the four bands removes mainly very faint, unresolved galaxies.

\begin{figure}
   \centering
   \includegraphics[width=\columnwidth]{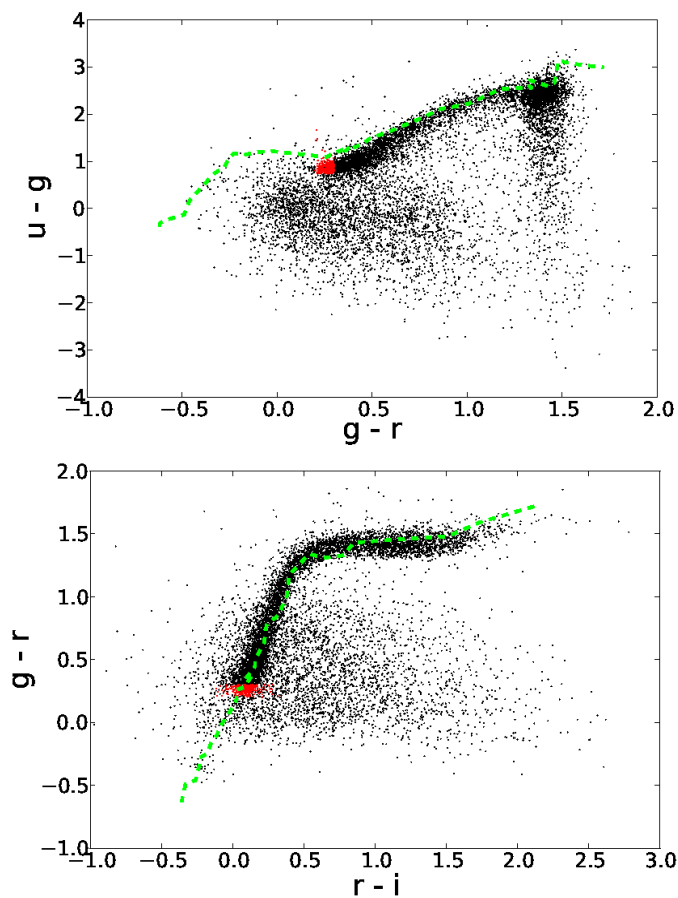}
      \caption{Colour-colour diagrams (CCDs) corresponding to the fields in group A 
               (pointings marked as light cyan circles in Figure~\ref{fig:mapPointings}). 
               The sources in the $ugri$ catalogues (black) and the subset of near-MSTO 
               stars (red) have been calibrated to SDSS using DR8 stellar photometry. 
               The main sequence stellar loci (green dashed lines) are the ones given 
               in Tables 3 and 4 of \citet{covey07}. 
               Quasars and white dwarf-M dwarf pairs are abundant in the $u-g<1$, 
               $-0.3<g-r<0.7$ space. 
               }
         \label{fig:slrCCD}
   \end{figure}

   We correct for interstellar extinction using the maps from \citet{schlegel98dustmaps} 
   and transform the magnitudes in the 
   $ugri$-stellar catalogues from the CFHT to the SDSS photometric system. For this we use 
   the equations on the Canadian Astronomy Data Center MegaCam website\footnotemark[1] 
   %which we subtract and substitute in order to obtain the reversed transformations:
   \begin{eqnarray}
      u_{MegaCam} &=& u_{SDSS} - 0.241\cdot(u_{SDSS} - g_{SDSS})  \\
      g_{MegaCam} &=& g_{SDSS} - 0.153\cdot(g_{SDSS} - r_{SDSS})  \\
      r_{MegaCam} &=& r_{SDSS} - 0.024\cdot(g_{SDSS} - r_{SDSS})  \\
      i_{MegaCam} &=& i_{SDSS} - 0.003\cdot(r_{SDSS} - i_{SDSS})  \,
   \end{eqnarray} 
   and invert them to turn our measurements into SDSS magnitudes. Subsequently we calibrate 
   each field directly to SDSS using stellar photometry from DR8. The resulting photometry 
   matches the colour-colour stellar loci of \citet{covey07} as shown in Figure~\ref{fig:slrCCD}. 
   Unless explicitly stated otherwise, all magnitudes in this paper are expressed in the 
   SDSS system.
   \footnotetext[1]{www2.cadc-ccda.hia-iha.nrc-cnrc.gc.ca/megapipe/docs/filters.html}
   
   In order to reduce the noise when analysing the radial stellar density distribution of the halo, 
   we combine the catalogues from nearby pointings, grouping them according to their 
   position in the sky. This step is important because of the nature of our survey, which is composed 
   of relatively small, scattered fields of view. 
   %Grouping the individual pointings increases the statistics of the stellar density measurements 
   %used for our analysis. 
   We use a friends-of-friends (FoF) algorithm to group the different pointings. 
   We request two friends not to be apart by more than 
   20 degrees, and in a few cases we clean or split a resulting group (red pentagons or blue 
   and orange triangles in Figure~\ref{fig:mapPointings}) or combine others (purple diamonds) 
   to account for the positions of the galactic disk or major halo substructures. Because the 
   different pointings in our surveys have different completeness limits, these 
   grouped or combined catalogues --which we name A,B,C,... H-- are finally filtered to meet 
   the completeness magnitude threshold of their most restrictive contributor\footnotemark[2].
   \footnotetext[2]{To determine the completeness limit of each field of view, we fit its 
   magnitude distribution to a gaussian --representing the population of faint galaxies-- and 
   another variable function --representing the stellar distribution along the whole magnitude 
   range--. We choose as the completeness limit either the transition point between the two 
   distributions (the valley) or, if instead there is a plateau, the turning point of the 
   whole distribution (the knee).}

   % Table 1: grouped fields of view - summary
   %*********************************************
   \begin{table*}
   \caption{Groups of pointings as shown in Figures~\ref{fig:mapPointings}, \ref{fig:densProfs}, 
   \ref{fig:fits_0.2mag} and \ref{fig:mapStreams}. 
   The table shows the central coordinates for each group, the number of individual fields 
   of view contributing to it, its total area and the stellar completeness limit in the r band.
            }                      
   \label{table:groups_of_view}                   % is used to refer this table in the text
   \centering                                                    % used for centering table
   \begin{tabular}{l r r r r c c c}                          % centered columns (4 columns)
   \hline%\hline                                          % inserts double horizontal lines
                                                                            % table heading 
   Group & $RA$ (deg) & $Dec$ (deg) & $l$ (deg) & $b$ (deg) & $n_{fields}$ & $\Sigma$ (deg$^2$) 
   & $ mag_{lim,r,*} $ (mag) \\
   \hline                                                  % inserts single horizontal line
                                                                        % body of the table
   A & 160.654338 & 43.98310 & 171.335811 & 59.15040 & 8 & 5.60 & 22.8 \\ 
   B & 231.593130 & 29.13513 &  45.577138 & 55.93598 & 5 & 3.98 & 22.7 \\ 
   C & 229.347757 &  6.91624 &   9.425402 & 49.92775 & 4 & 3.44 & 24.1 \\ 
   D & 210.062933 & 51.67173 &  99.735627 & 62.24580 & 2 & 0.64 & 23.4 \\ 
   E & 121.918411 & 41.20348 & 179.233500 & 31.26694 & 5 & 2.73 & 22.7 \\ 
   F & 342.735895 & 17.09581 &  86.019738 & -36.99391 & 3 & 2.17 & 23.2 \\ 
   G & 157.028363 & 17.15674 & 222.142793 & 55.48268 & 3 & 2.02 & 23.1 \\ 
   H & 220.659749 &  2.00187 & 354.337092 & 53.38989 & 3 & 2.04 & 24.2 \\ 
   \hline                                                             %inserts single line
   \end{tabular}
   \end{table*}

   % Figure: CMDs for gr, ugri and difference
   %*******************************************

   % Figure: SLR colour-colour calibration
   %*****************************************

   % Figure: Abs. mag. and metallicity of nearMSTOstars
   %****************************************************
   \begin{figure}
   \centering
   \includegraphics[width=\columnwidth]{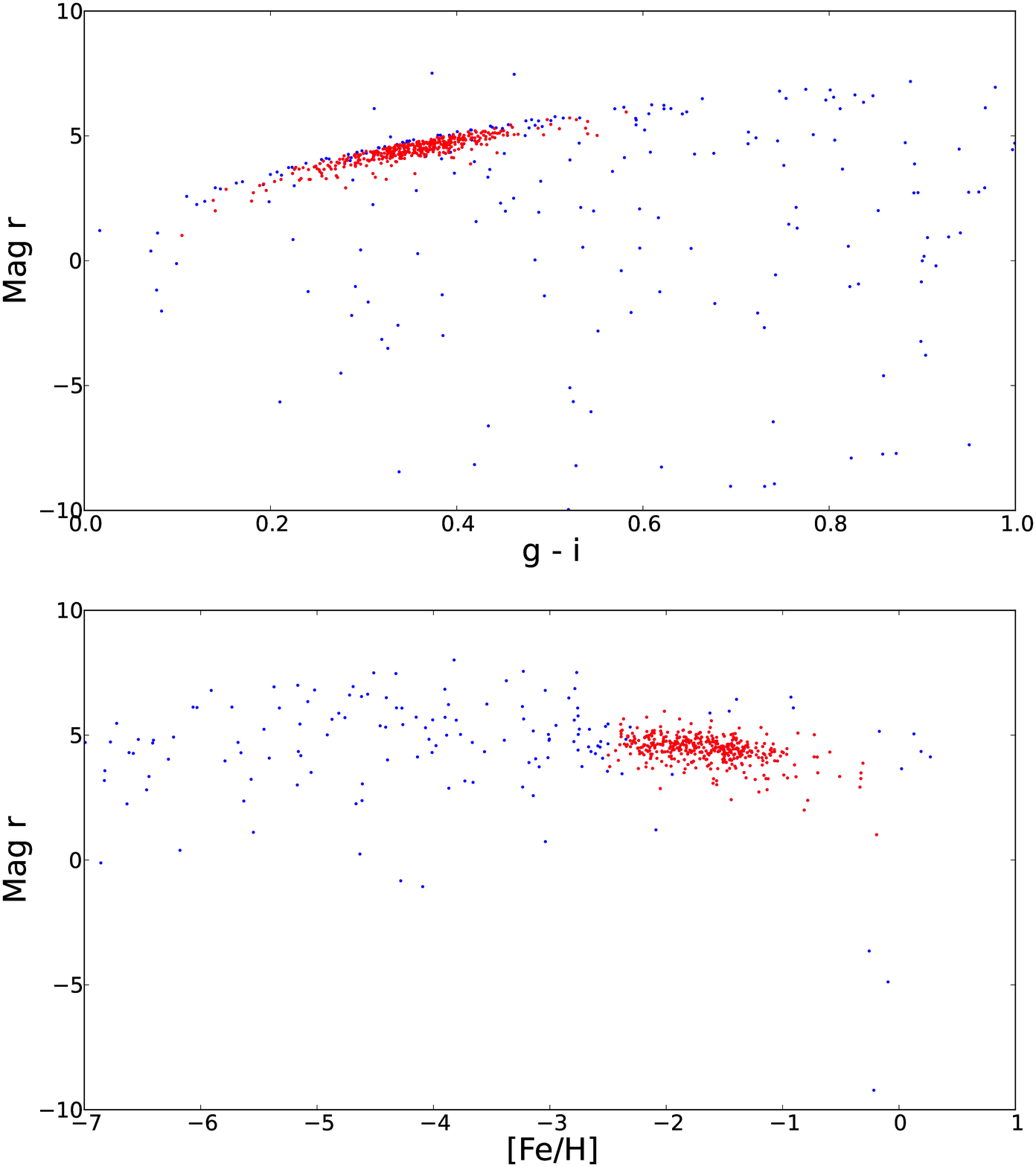}
      \caption{Estimated absolute magnitude in the $r$ band ($M_r$) and estimated metallicity 
               ($[Fe/H]$) for group A for the sources typically considered as halo stars (blue) 
               and those that we have selected as near-MSTO stars (red). The sources selected as halo 
               members meet $0.2<g-r<0.3$ and $g,r,i>17$. The subset of near-MSTO stars, 
               additionally meets $M_r>-2$, $-2.5 \leq [Fe/H] \leq 0$ and $0.1 < g-i < 0.6$. 
               }
         \label{fig:absMag_Metal}
   \end{figure}

%%%%%%%%%%%%%%%%%%%%%%%%%%%%%%%%%%%%%%%%%%%%%%%%%%%%%%%%%%%%%%%%%%%%%%%%%%%%%%%%%%%%%%%%%%%%
%  _________________________________________________________________________________________
% 
%  3. DENSITY PROFILES
%  _________________________________________________________________________________________
%
%%%%%%%%%%%%%%%%%%%%%%%%%%%%%%%%%%%%%%%%%%%%%%%%%%%%%%%%%%%%%%%%%%%%%%%%%%%%%%%%%%%%%%%%%%%%

\section{Stellar radial density profiles}\label{results}

   % ---------------------------------------------------------------------------------------
   % 3.1 NEAR-MSTO STARS - SELECTION
   % ---------------------------------------------------------------------------------------

   \subsection{Star selection and construction of the radial stellar density profiles} 
   \label{subsec:nMSTOselec}
   
   The coordinates and the completeness limits of the groups are given in 
   Table~\ref{table:groups_of_view}. 
   %The completeness limit is dependent on the limiting magnitudes of the four different bands, 
   %and thus on the varying observing conditions across each group of pointings. 
   %Among them there is a subset of halo near-MSTO stars that stretches out to $80-100kpc$   
   %from the galactic centre, depending on the line of sight. We use these near-MSTO stars to 
   %measure their number density distribution as a function of position in the Galaxy. 
   We use halo main sequence turnoff stars in our fields as tracer of the stellar halo: at 
   the completeness limits of the data such stars can be identified as far out as $60$ kpc 
   from the Galactic centre.
   We fit several Galactic stellar distribution models to %it and we derive from them 
   these density profiles and derive a number of structural parameters for the stellar halo. 
   Previous works have already used main sequence turnoff point (MSTO) stars, 
   near-MSTO stars, BHB and blue stragglers of type A and RRLyrae as stellar tracers for the 
   Galactic stellar halo. %in different data surveys. 
   We compare and discuss our findings to theirs in section~\ref{subsec:compData}. 
   
   In order to select the near main sequence turnoff stars we make use of two empirical 
   photometric variables. The ratio $[Fe/H]$ is calculated following the photometric metallicity relation 
   by \citet{bond10}, and the absolute magnitude $M_r$ is calculated following the photometric 
   parallax relation from \citet{ivezic08}:

   \begin{multline}
      [Fe/H] = -13.13 + 14.09x + 28.04y - 5.51xy -5.90x^2 \\
       - 58.68y^2 + 9.14x^2y - 20.61xy^2 + 58.20y^3, \, \, \, \, \, \, \, \,  
   \end{multline} 
   where $x=u-g$ and $y=g-r$. This relation is valid in the $g-i < 0.6$ and 
   $-2.5 \leq [Fe/H] \leq 0$ range, which is compatible with the regime 
   of our near-MSTO star selection. 

   \begin{multline}
      M_r = -0.56 + 14.32z - 12.97z^2 + 6.127z^3 -1.267z^4 \\
       + 0.0967z^5 -1.11[Fe/H] - 0.18[Fe/H]^2, \, \, \, \, \, \, \, \,  \, \, \, \, \, 
   \end{multline}
   where $z=g-i$. The tested validity regime of this equation encompasses the $0.2<g-i<1.0$ 
   range, meaning that the absolute brightnesses of our near-MSTO stars have been properly 
   estimated. We extrapolate the relation for the $0.1<g-i<0.2$ range, which is justified 
   owing to the smooth and slow change of $M_r$ with $z$.
   
   We select the halo near-MSTO stars by requiring 
   \begin{eqnarray}
      0.2 < g-r < 0.3 \,; \label{constr:g-r} \\ 
      g,r,i > 17 \,; \label{constr:gri} \\
      0.1 < g-i < 0.6 \,; \label{constr:g-i} \\  
      5.0 > M_r > -2 \,; \label{constr:absMag} \\ 
      -2.5 \leq [Fe/H] \leq 0 \label{constr:metal} \,. % \mathrm{dex}
   \end{eqnarray} 
   
   The first two restrictions (\ref{constr:g-r} and \ref{constr:gri}) retrieve stars typically 
   associated with the halo, in particular distant main sequence F stars (see Table 3 from 
   \citet{covey07}). This selection however, can be significantly contaminated by quasars and 
   white dwarf-M dwarf pairs, which are abundant in (but not restricted to) the $-0.2<g-r<0.3$ 
   range (see Figure~\ref{fig:slrCCD}). To reduce the presence of these interlopers and 
   select the bulk of the F stars population, we apply restrictions~\ref{constr:g-i} (based on 
   Table 4 in \citet{covey07}) and \ref{constr:absMag}. %The lower limit in 
   %constraint~\ref{constr:absMag} corresponds to the absolute brightness of the tip of the 
   %M-giant stars, and thus removes any stellar or stellar-like source brighter than that 
   %(as far as their absolute brightness has been properly estimated), while preserving 
   %the main sequence F stars. \textbf{Its upper limit guarantees the completeness of 
   %these stars at our distance limit. }
   Constraint~\ref{constr:metal} ensures that the final sources are at most 
   as metal rich as the Sun (to account for possible contributions from metal-rich 
   satellites) and not more metal-poor than 0.003 times the Sun. 
   
   The decrease in interlopers attained by applying restrictions~\ref{constr:g-i}, 
   \ref{constr:absMag}, and \ref{constr:metal} compared to only applying restrictions 
   \ref{constr:g-r} and \ref{constr:gri} is illustrated in Figure~\ref{fig:slrCCD}, where 
   the red dots indicate the final selection of halo near-MSTO stars and the black dots 
   represent the whole catalogue of star-like sources. It is clear that the 
   final selection of near-MSTO stars does not span the whole range of sources encompassed 
   between $g-r=0.2$ and $g-r=0.3$. The effect of the $[Fe/H]$ and $M_r$ selection is 
   further illustrated in Figure~\ref{fig:absMag_Metal}. 
   
   Using the estimated absolute brightness, we calculate the distance modulus and the 
   heliocentric distance for all the near-MSTO stars. We define distance modulus bins of 
   size $\Delta\mu=0.2$ mag and $\Delta\mu=0.4$ mag, and count the number of near-MSTO stars 
   per bin for each group of fields (A,B,C,...). The choice of distance bins is 
   motivated by a compromise between maximising the radial distance resolution and 
   minimising the Poisson noise in the stellar number counts. We test this compromise by 
   exploring two distance modulus bin sizes, which correspond to distance bin sizes of the 
   order of $10^2$ pc and $10$ kpc, respectively.

   We then calculate the number density per bin and its uncertainty as follows:
   \begin{eqnarray}
      \rho_{l,b,D} &=& \frac{N_{l,b,\Delta\mu}}{0.2\cdot ln(10)\cdot D_{hC}^3\cdot 
                    \Delta\Omega\cdot \Delta\mu} \label{eq:dens} \,, \\ 
      E_{\rho} &=& \sqrt{(\frac{\rho}{\sqrt{N}})^2 + (\frac{\rho}{\sqrt{n_{fields}}})^2}     
                    \label{eq:err_dens} \,, 
   \end{eqnarray} 
   where $\Delta\Omega$ is the area covered by each group, $D_{hC}$ is the heliocentric 
   distance, $l$ and $b$ are the galactic coordinates and $N_{l,b,\Delta\mu}$ is the number 
   of stars per bin in a given direction of the sky. Particularly, 
   \begin{eqnarray} 
      \Delta\Omega &=& \frac{4\pi}{41253} \Sigma\mathrm{(deg^2)} \label{eq:beamArea} \, %\\
      %E_{\Delta\Omega} &=& \frac{4\pi}{41253} \cdot E_{\Sigma} \label{eq:errArea} \,; 
   \end{eqnarray} 
   and the area of each group ($\Sigma$) depends on the individual area of each field 
   contributing to it (Table~\ref{table:groups_of_view}). 
   %In equation~\ref{eq:err_dens} 
   %we assume that the systematic bias in our photometric uncertainties is negligible, and 
   %therefore the poissonian noise in the number of stars per bin and the area of each group 
   %suffice to weight the uncertainty in the number density. 
   
\begin{figure*}
   \centering
   \includegraphics[width=\textwidth]{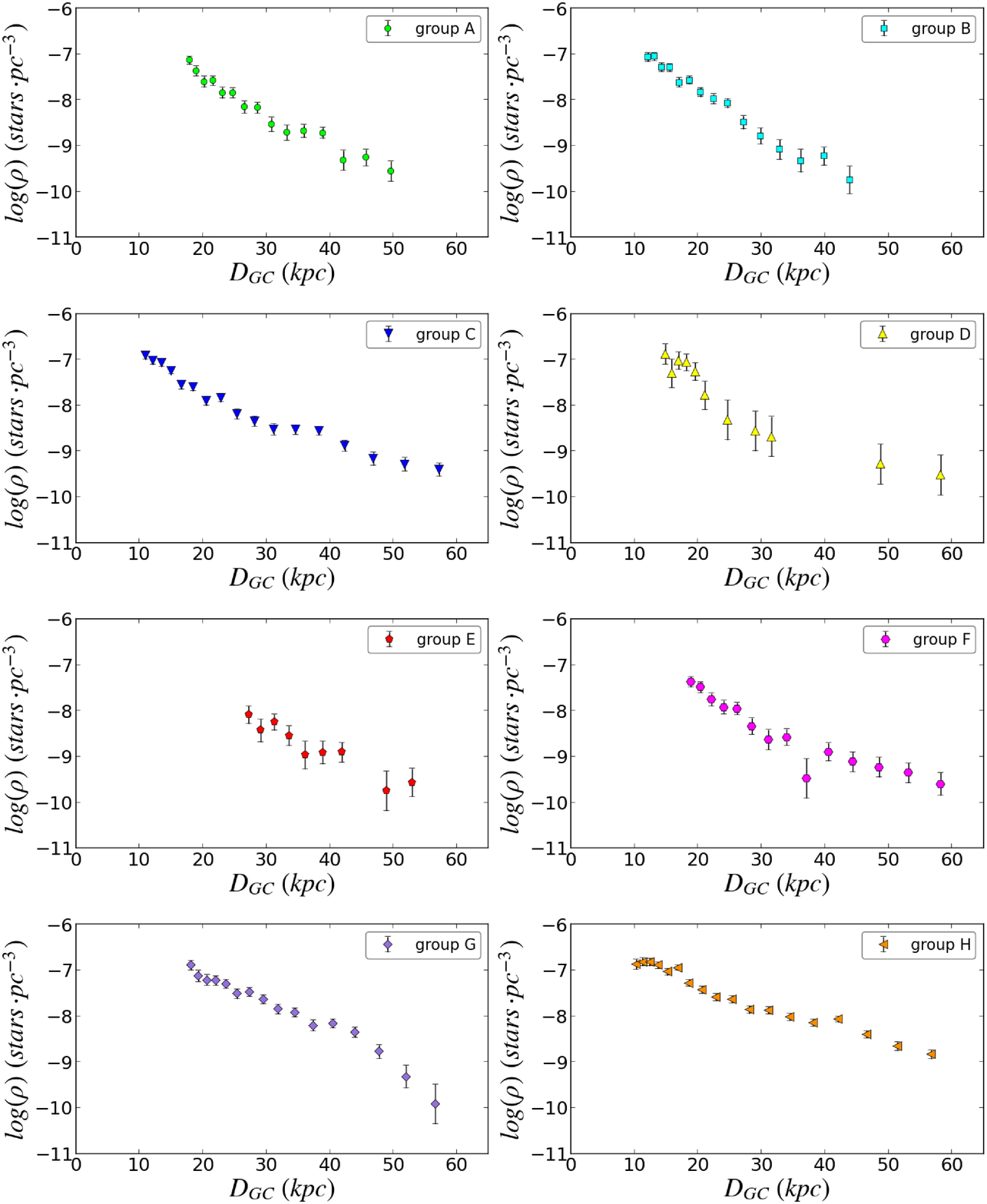}
      \caption{Logarithmic stellar density profiles versus distance for the near Main Sequence 
               turnoff point stars (near-MSTO) from the fields in groups A (green circles), B 
               (cyan squares), C (blue downward triangles), D (yellow upward triangles), E (red 
               pentagons), F (pink hexagons), G (purple diamonds) and H (orange leftward 
               triangles). Their symbols match those in Figure~\ref{fig:mapPointings}.
               }
         \label{fig:densProfs}
   \end{figure*}

   The results for these number density calculations can be seen in Figure~\ref{fig:densProfs}, 
   where we plot the logarithmic number density against the galactocentric distance\footnotemark[3], 
   $R_{GC}$, for each group (or line of sight). For this and the subsequent analysis, we only 
   consider bins with $R_{GC}>5\mathrm{kpc}$, $|z|>10$ kpc (to avoid the inner regions of the 
   Galaxy) and a distance modulus of $\mu \leq mag_{lim} -4.5$ (to guarantee a 
   complete sample of the faintest near-MSTO stars\footnotemark[4]). 
   \footnotetext[3]{
      \begin{eqnarray*}
         R_{GC} = \sqrt{R_{GC}^2 + z^2} \, %\\
      \end{eqnarray*}
      where $R_{GC}$ and $z$ are the radial and vertical coordinates on the cylindrical 
      galactocentric reference system.
   }
   \footnotetext[4]{This constraint guarantees that there are no distance completeness issues 
    due to our specific type of stellar tracers and due to the different depths of our fields. 
    The only subset affected by incompleteness is that of $mag_{lim} -5.0 < \mu < mag_{lim} -4.5$ 
    for the stars in the $4.5 < M_r < 5.0$ range; and its average loss  is of $20\%$ over the 
    total number of near-MSTO stars ($-2.0 <M_r<5.0$) in the same distance range. Several tests 
    on different upper distance thresholds for the density profiles show that the distance modulus 
    constraint of $\mu \leq mag_{lim} -4.5$ is enough to guarantee that all the lines of sight 
    contribute robust density measurements at the furthest distances and that the incompleteness 
    in $mag_{lim} -5.0 < \mu < mag_{lim} -4.5$ for the $4.5 < M_r < 5.0$ near-MSTO stars has no 
    statistically significant effect on the best fit parameters. 
    %The average loss due to completeness beyond  $\mu > mag_{lim} -5.0$ for the stars in the 
    %$4.5<M_r<5.0$ range is a $20\%$ over the total number of near-MSTO stars at 
    %$mag_{lim} -5.0 < \mu < mag_{lim} -4.5$, and this has no strong influence on the outer halo parameters. 
    }
   
   % Figure: Density profiles
   %****************************
   
   Figure~\ref{fig:densProfs} shows that the density profiles decrease quite smoothly for 
   $40-60$ kiloparsecs and %start to drop to 1 or 0 star counts per bin at around 60-90 kpc 
   for most of the lines of sight. %The density profile for group D is particularly affected 
   %by large errorbars and a shallower view. This can be explained by its relatively small area, 
   %composed of just two rather narrow pointings. 

   % ---------------------------------------------------------------------------------------
   % 3.2 Fitting procedure
   % ---------------------------------------------------------------------------------------
   
   \subsection{Fitting procedure}\label{subsec:fittings}
   
   We fit several models of the Galactic stellar number density distribution to the data, 
   ranging from a basic axisymmetric power law to more complex models with triaxiality and 
   a break in the power law. 
   The models take the following mathematical forms, with $x$, $y$, and $z$ being the cartesian 
   galactocentric coordinates with the Sun at (8,0,0) kpc \citep{malkin12}:
   \begin{itemize}
   %\item[-] Spherical model:
   %   \begin{eqnarray}
   %   \rho(x,y,z) &=& \rho_0\cdot \left(x^2 + y^2 + z^2\right)\, ^{n/2} 
   %                   \label{eq:spherical} \,.
   %   \end{eqnarray} 
   %
   \item[-] Axisymmetric model
      \begin{eqnarray}
      \rho(x,y,z) &=& \rho_0\cdot \left(x^2 + y^2 + \frac{z^2}{q}\right)\, ^{n/2} 
                      \label{eq:axisym} \,, 
      \end{eqnarray} 
      where $q=c/a$ is the polar axis ratio or the oblateness of the halo;
   
   \item[-] Triaxial model
      \begin{eqnarray}
      \rho(x,y,z) &=& \rho_0\cdot \left(x^2 + \frac{y^2}{w} + \frac{z^2}{q}\right)\, ^{n/2} 
                      \label{eq:triaxial} \,,  
      \end{eqnarray} 
      where $w=b/a$ is the ratio between the axes in the Galactic plane;
      
   \item[-] Broken power law, with varying power index at $R_{break}$
      \begin{eqnarray}
       \rho(x,y,z)=\left\{
                 \begin{array}{ll}
                  \rho_0\cdot \left(R_{ellip}\right)\, ^{n_{in}}\,, \, \, \, \, \, 
                       \, \, \, \, \, \,  \, \, \, \, \, \,  \, \, \, \, \, \, R_{ellip}<R_{break} 
                      \label{eq:simplebroken}  \\
                  \rho_0\cdot \left(R_{ellip}\right)\, ^{n_{out}} \cdot R_{break}^{n_{in}-n_{out}} 
                      \,, \, \, R_{ellip}\geq R_{break} 
                \end{array}
              \right. \\
      R_{ellip} = \left(x^2 + y^2 + \frac{z^2}{q}\right)\, ^{1/2}  \,; \notag
      \end{eqnarray}
   
   \item[-] Broken power law, with varying power index and oblateness at $R_{break}$
      \begin{eqnarray}
       \rho(x,y,z)=\left\{
                 \begin{array}{ll}
                  \rho_0\cdot \left(x^2 + y^2 + \frac{z^2}{q_{in}}\right)\, ^{n_{in}/2}\,, 
                      \, \, R_{GC}\leq R_{break} 
                      \label{eq:broken}  \\
                  \rho_0\cdot \left(x^2 + y^2 + \frac{z^2}{q_{out}}\right)\, ^{n_{out}/2}\,, 
                      \, \, R_{GC}>R_{break} \,,  \notag
                \end{array}
              \right. \\
      \end{eqnarray}
      where the inner power law is fit to data that meets $R_{GC}\leq R_{break}$ and the outer power 
      law is applied to data that meets $R_{GC}>R_{break}$.
   
   \end{itemize}

   We fit all these models to the data using the "curve-fit" method from Scipy.optimize, which uses 
   the Levenberg-Marquardt algorithm for non-linear least squares fitting. The objective function 
   takes the form of a $\chi^2$, and we also calculate a reduced $\chi^2$ for analysis purposes, 
   \begin{eqnarray}
      \chi^2 &=& \sum_{i=1}^{N_{data}} \left(\frac{\rho_{data,i}-\rho_{model,i}}{E_{\rho,i}}\right)^2 \,, 
                 \label{eq:chi2} \\ 
      \chi_{red}^2 &=& \frac{\chi^2}{N_{data}-N_{params}}     
                    \label{eq:red_chi2} \,, 
   \end{eqnarray}
   where $N_{data}$ and $N_{params}$ are the number of data points and the number of free parameters, 
   respectively. 

   The influence of the photometric uncertainties on the density profiles and the best fit parameters 
   is evaluated through a set of Monte Carlo simulations that randomly modify the $g$,$r$,$i$,$u$ 
   magnitudes of each star within the limits of the photometric uncertainties. Through this method 
   we find that the variation of the Monte Carlo best fit parameters aligns with the uncertainties of 
   our best fit parameters (derived from the second derivative of the fits by the "curve-fit" method). 
   The centre of these variations is within $1\sigma$ of our direct findings.

   We fit all models to four data sets: with and without [known] substructures and binned in $0.2$ 
   and $0.4$ magnitude cells. In this way we can check the robustness of our results to different 
   binning options and we are able to compare what would be the effect of substructure on our 
   understanding of the smooth halo if we were to ignore it or unable to recognize it as such. 
   Specifically, we cut the distance bins at $R_{GC}<25$ kpc in group E to avoid contributions 
   by the structures in the direction of the galactic anticentre (the Monoceros ring, the 
   Anticentre Structure and the Eastern Band Structure), the distance bins within $15<D_{hC}<40$ kpc 
   in group G to avoid contributions by the Sagittarius stream, and the distance bins within 
   $20\, \mathrm{kpc}<D_{hC}<60$ kpc in group H to avoid contributions again by the Sagittarius 
   stream.

   % ---------------------------------------------------------------------------------------
   % 3.3 Results
   % ---------------------------------------------------------------------------------------
   
   \subsection{Results}\label{subsec:results}
   
   The best fit parameters for each model resulting from fitting these four data sets are 
   summarized in Tables~\ref{table:params_0.2mag_nosubstr} to \ref{table:params_0.4mag_alldata}. 
   Table~\ref{table:params_0.2mag_nosubstr} contains the results of fitting the $\Delta\mu=0.2$ mag 
   binned data excluding regions with substructure, whereas Table~\ref{table:params_0.2mag_alldata} 
   contains the results of fitting to all the $0.2$ mag bins. 
   Similarly Table~\ref{table:params_0.4mag_nosubstr} covers the fits to $\Delta\mu=0.4$ mag data 
   without substructure bins, and Table~\ref{table:params_0.4mag_alldata}, to all $0.4$ mag bins. 
   The reduced $\chi^2$ and the initial parameters have also been recorded in these tables.

   \begin{sidewaystable*}
   
   % Table 2a: best fit parameters (with boundaries, without substructure), 0.2mag bins
   %*************************************************************************************
   \caption{Best fit parameters for the four different Galactic stellar distribution models 
            resulting from removing the data that is affected by known halo substructures 
            (the Sagittarius stream and the anticentre substructures). 
            For the fitting, the data has been binned in $0.2$ mag distance modulus cells.
            %Same as in table~\ref{table:bound_params} but this time letting the possible 
            %parameter values evolve without constraints.
            }                      
   \label{table:params_0.2mag_nosubstr}             % is used to refer this table in the text
   %\centering                                                    % used for centering table
   \begin{tabular}{l r c c c c c c c c c}                     % centered columns (4 columns)
   \hline%\hline                                          % inserts double horizontal lines
                                                                            % table heading 
    Model & $\chi_{red}^2$ & $\rho_0\, \mathrm{(pc^{-3})}\cdot 10^{-3}$ 
    & $R_{break}\, \mathrm{(kpc)}$ & $n$ & $n_{in}$ & $n_{out}$ & $q$ & $q_{in}$ & $q_{out}$ 
    & $w$ \\
   \hline                                                  % inserts single horizontal line
   \smallskip                                                           % body of the table
   %spherical & 870.59 & 3.74 & $5.06\cdot 10^{-4}$ & -3.45 & -- & -- 
   %           & -- & -- & -- & -- & -- & -- \\ 
   axisymmetric & 1.90 & $14\pm6$ & -- & $-4.31\pm0.09$ & -- & -- 
               & $0.62\pm0.06$ & -- & -- & -- \\ 
   triaxial & 1.86 & $14\pm6$ & -- & $-4.28\pm0.09$ & -- & -- 
            & $0.60\pm0.06$ & -- & -- & $0.75\pm0.09$ \\ 
   broken p.l.$_n$ & 1.52 & $0.071\pm0.003$ & $19.0\pm0.5$ 
                        & -- & $-2.40\pm0.05$ & $-4.80\pm0.05$ 
                               & $0.60\pm0.03$ & -- & -- & -- \\ 
   broken p.l.$_{n,\, q}$ & 1.99,1.51 & $1\pm3$ 
                               & $19_{fixed}$
                               & -- & $-3.3\pm0.6$ & $-4.9\pm0.2$ 
                               & -- & $0.5\pm0.2$ & $0.68\pm0.07$ & -- \\ 
   \hline                                                             %inserts single line
   initial parameters & -- & $0.001$ & 40.0 & -3.00 & -3.00 & -3.50 
                               & 0.70 & 0.70 & 0.8  & 1.00 \\ 
   \hline    
   \end{tabular}
 
   \bigskip\bigskip  % provide some separation between the two tables
   
   % Table 2b: best fit parameters (with boundaries, with substructure), 0.2mag bins
   %**********************************************************************************
   \caption{Same as in Table~\ref{table:params_0.2mag_nosubstr} but this time fitting all the 
            available data (including those regions containing stellar counts from known 
            substructures and detected overdensities).
            }                      
   \label{table:params_0.2mag_alldata}             % is used to refer this table in the text
   %\centering                                                    % used for centering table
   \begin{tabular}{l r c c c c c c c c c}                     % centered columns (4 columns)
   \hline%\hline                                           % inserts double horizontal lines
                                                                            % table heading 
    Model & $\chi_{red}^2$ & $\rho_0\, \mathrm{(pc^{-3})}\cdot 10^{-3}$ 
    & $R_{break}\, \mathrm{(kpc)}$ & $n$ & $n_{in}$ & $n_{out}$ & $q$ & $q_{in}$ & $q_{out}$ 
    & $w$ \\
   \hline                                                  % inserts single horizontal line
   \smallskip                                                           % body of the table
   %spherical & 1465.47 & 5.12 & rho & n & -- & -- & -- & -- & -- & -- & -- & -- \\ 
   axisymmetric & 4.71 & $8\pm3$ & -- & $-4.15\pm0.08$ & -- & -- 
               & $0.69\pm0.06$ & -- & -- &  --  \\ 
   triaxial & 4.59 & $7\pm2$ & -- & $-4.07\pm0.08$ & -- & -- 
            & $0.68\pm0.06$ & -- & -- & $0.59\pm0.07$ \\ 
   broken p.l.$_n$ & 4.24 & $0.17\pm0.01$ & $21.0\pm0.5$ & -- & $-2.80\pm0.05$ & $-4.80\pm0.05$ 
                        & $0.70\pm0.03$ & -- & -- & -- \\ 
   broken p.l.$_{n,\, q}$ & 3.36,4.79 & $1\pm2$ & $21_{fixed}$ 
                          & -- & $-3.3\pm0.4$ & $-5.0\pm0.2$ 
                               & -- & $0.5\pm0.2$ & $0.79\pm0.08$  & -- \\ 
   \hline                                                             %inserts single line
   initial parameters & -- & $0.001$ & 40.0 & -3.00 & -3.00 & -3.50 
                               & 0.70 & 0.70 & 0.8  & 1.00 \\ 
   \hline 
   \end{tabular}
   
   \bigskip\bigskip  % provide some separation between the two tables

   % Table 3a: best fit parameters (with boundaries, without substructure), 0.4mag bins
   %*************************************************************************************
   \caption{Same as in Table~\ref{table:params_0.2mag_nosubstr} but this time fitting the 
            data binned in $0.4$ mag distance modulus cells.
            }                      
   \label{table:params_0.4mag_nosubstr}            % is used to refer this table in the text
   %\centering                                                    % used for centering table
   \begin{tabular}{l r c c c c c c c c c}                     % centered columns (4 columns)
   \hline%\hline                                          % inserts double horizontal lines
                                                                            % table heading 
    Model & $\chi_{red}^2$ & $\rho_0\, \mathrm{(pc^{-3})}\cdot 10^{-3}$ 
    & $R_{break}\, \mathrm{(kpc)}$ & $n$ & $n_{in}$ & $n_{out}$ & $q$ & $q_{in}$ & $q_{out}$ 
    & $w$ \\
   \hline                                                  % inserts single horizontal line
   \smallskip                                                           % body of the table
   %spherical & 870.59 & 3.74 & $5.06\cdot 10^{-4}$ & -3.45 & -- & -- 
   %           & -- & -- & -- & -- & -- & -- \\ 
   axisymmetric & 3.89 & $12\pm4$ & -- & $-4.26\pm0.08$ & -- & -- 
               & $0.60\pm0.05$ & -- & -- & -- \\ 
   triaxial & 3.97 & $12\pm5$ & -- & $-4.25\pm0.08$ & -- & -- 
            & $0.60\pm0.06$ & -- & -- & $0.9\pm0.1$ \\ 
   broken p.l.$_n$ & 2.61 & $0.11\pm0.01$ & $20.0\pm0.5$ & -- & $-2.60\pm0.05$ & $-4.90\pm0.05$ 
                               & $0.65\pm0.03$ & -- & -- & -- \\ 
   broken p.l.$_{n,\, q}$ & 4.95,2.34 & $1\pm1$ & $20_{fixed}$ & -- & $-3.2\pm0.4$ 
                          & $-5.0\pm0.3$ %\tiny{unconstrained} 
                          & -- & $0.5\pm0.2$ & $0.67\pm0.08$ & -- \\ 
   \hline                                                             %inserts single line
   initial parameters & -- & $0.001$ & 40.0 & -3.00 & -3.00 & -3.50 
                               & 0.70 & 0.70 & 0.8  & 1.00 \\ 
   \hline 
   \end{tabular}
   
   \bigskip\bigskip  % provide some separation between the two tables
   
   % Table 3b: best fit parameters (with boundaries, with substructure), 0.4mag bins
   %**********************************************************************************
   \caption{Same as in Table~\ref{table:params_0.4mag_nosubstr} but this time fitting all the 
            available data (including those regions containing stellar counts from known 
            substructures and detected overdensities).
            }                      
   \label{table:params_0.4mag_alldata}             % is used to refer this table in the text
   %\centering                                                    % used for centering table
   \begin{tabular}{l r c c c c c c c c c}                     % centered columns (4 columns)
   \hline%\hline                                           % inserts double horizontal lines
   \smallskip                                                           % body of the table
    Model & $\chi_{red}^2$ & $\rho_0\, \mathrm{(pc^{-3})}\cdot 10^{-3}$ 
    & $R_{break}\, \mathrm{(kpc)}$ & $n$ & $n_{in}$ & $n_{out}$ & $q$ & $q_{in}$ & $q_{out}$ 
    & $w$ \\
   \hline                                                  % inserts single horizontal line
                                                                        % body of the table
   %spherical & 1465.47 & 5.12 & rho & n & -- & -- & -- & -- & -- & -- & -- & -- \\ 
   axisymmetric & 9.13 & $7\pm2$ & -- & $-4.10\pm0.07$ & -- & -- 
               & $0.66\pm0.05$ & -- & -- &  --   \\ 
   triaxial & 9.19 & $7\pm2$ & -- & $-4.07\pm0.07$ & -- & -- 
            & $0.65\pm0.06$ & -- & -- & $0.74\pm0.09$ \\ 
   broken p.l.$_n$ & 7.74 & $0.058\pm0.005$ & $20.0\pm0.05$ & -- & $-2.40\pm0.05$ & $-4.8\pm0.05$ 
                        & $0.70\pm0.03$ & -- & -- & -- \\ 
   broken p.l.$_{n,\, q}$ & 6.05,9.2 & $0.6\pm0.9$ & $20_{fixed}$ & -- & $-3.1\pm0.4$ 
                              & $-4.9\pm0.2$ 
                               & -- & $0.5\pm0.2$ & $0.74\pm0.07$ & -- \\ 
   \hline                                                             %inserts single line
   initial parameters & -- & $0.001$ & 40.0 & -3.00 & -3.00 & -3.50 
                               & 0.70 & 0.70 & 0.8  & 1.00 \\ 
   \hline 
   \end{tabular}
   \end{sidewaystable*}

   We compare the fitting results for the four different data sets recorded in 
   Tables~\ref{table:params_0.2mag_nosubstr} to \ref{table:params_0.4mag_alldata} and find 
   that the fits for which the substructure has been masked significantly outperform 
   those that have been allowed to fit all the available data. The difference on $\chi^2_{red}$
   for all these models and bin sizes is in every case at least a factor of $2.3$ or larger. 
   We find that allowing the models to fit data that contains substructure does not affect 
   largely most of the structural parameters (polar axis ratios are compatible within the 
   uncertainties and power law indices have close values) except that 
   it decreases the disk axis ratio $w$ by at least $15\%$, suggesting a strong departure 
   from the axisymmetric model that is not implicit in the filtered data sets. Henceforth we 
   will restrict the remaining discussion to the results derived from the cleanest data sets. 
   
\begin{sidewaysfigure*}
   \centering
    \subfloat[Fitted density profiles for the $0.2$ mag binned data.]{
             \includegraphics[width=12cm]{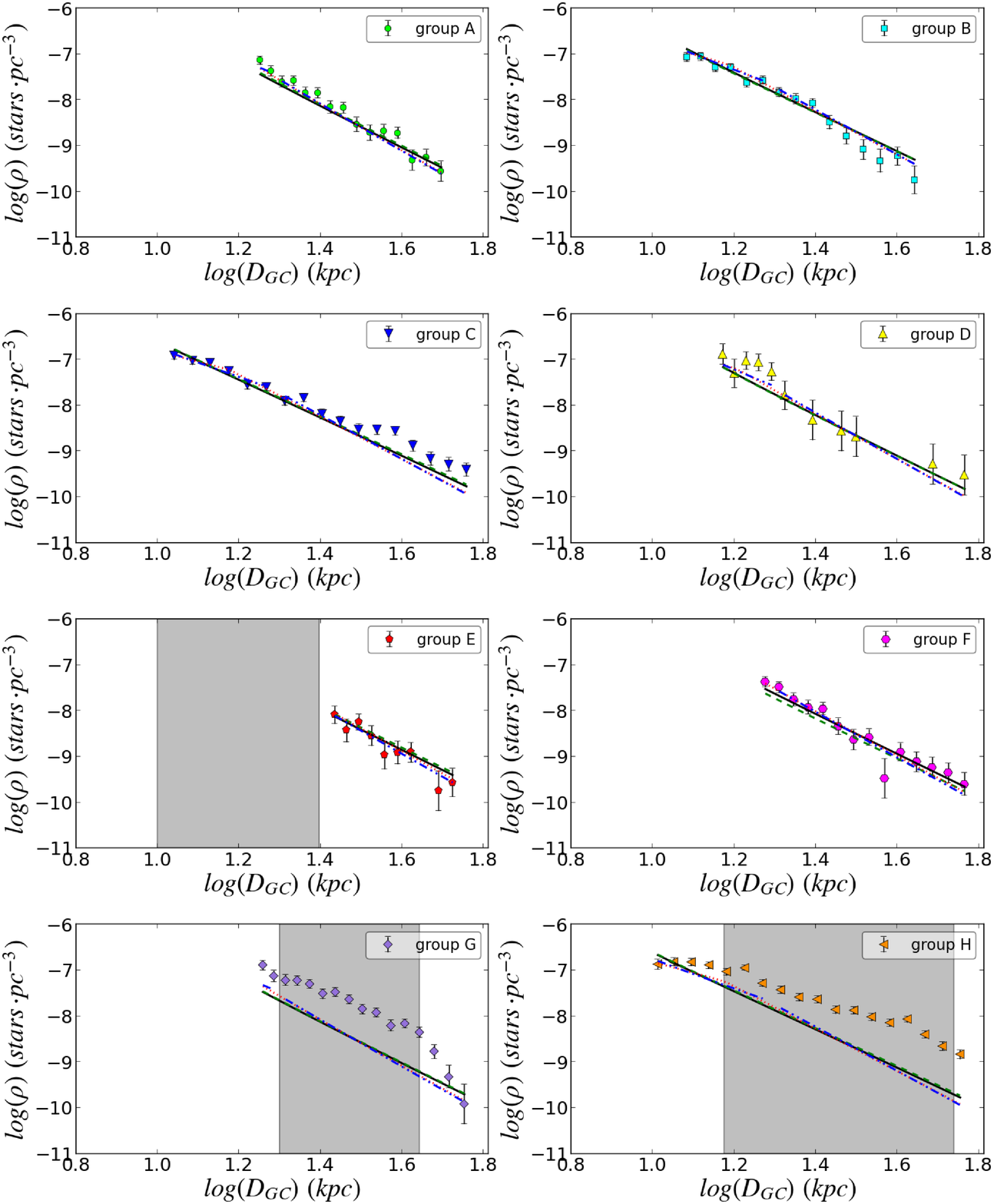}
             \label{fig:fitted_densProfs}
             }
    \subfloat[Data-to-model residuals for the $0.2$ mag binned data.]{
             \includegraphics[width=12cm]{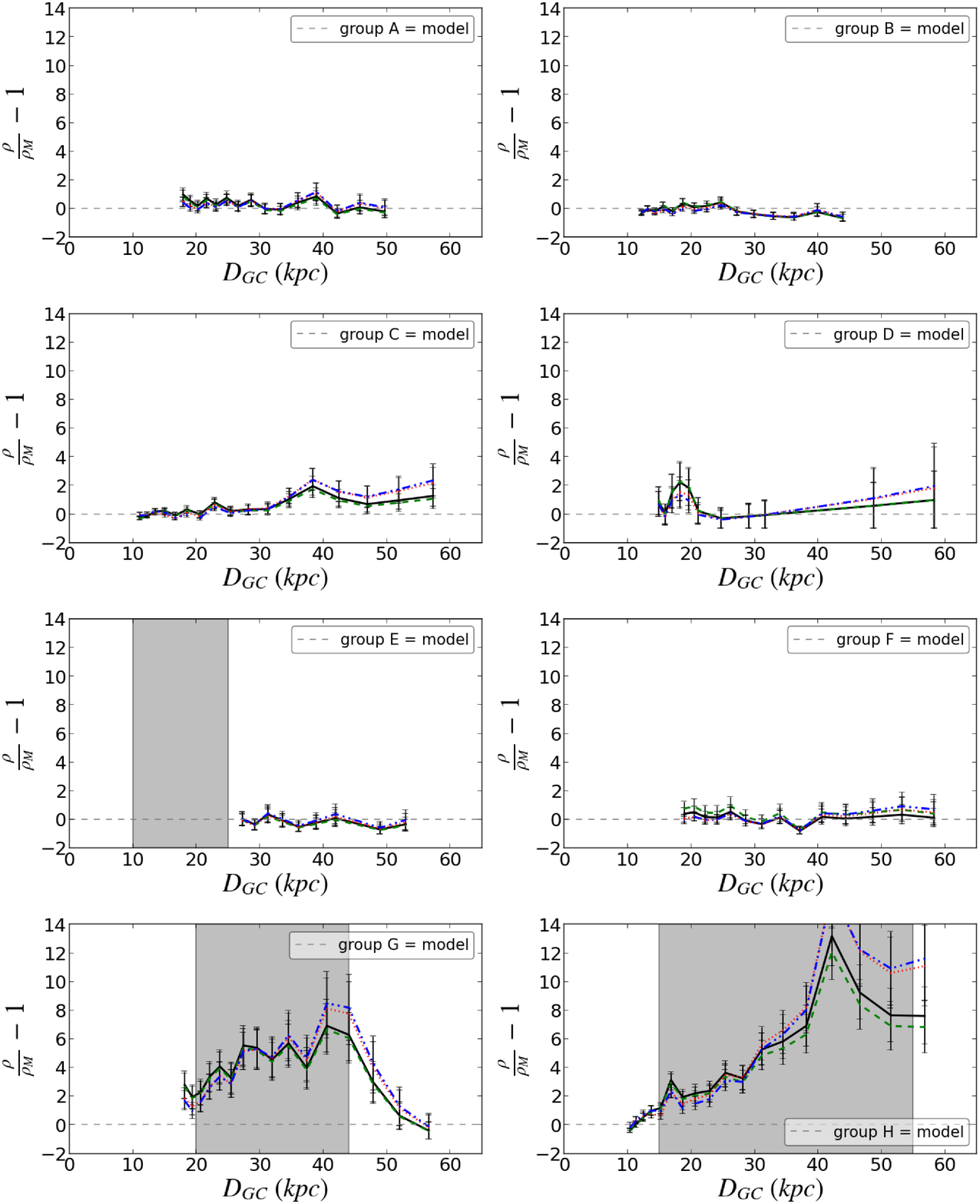}
             \label{fig:fits_residuals}
             }
    \caption{Left panels: density profiles in decimal logarithmic scale and the best fit models from 
               Table~\ref{table:params_0.2mag_nosubstr} (fitted to masked $0.2$ 
               binned data). 
               Right panels: Residuals between the data and the best fit models from 
               panel~\ref{fig:fitted_densProfs}. 
               The different lines represent 
               the axisymmetric (black solid line), the triaxial (green dashed line), 
               %the triaxial with free orientation of the x-y axes (red dashed-dotted line), 
               the broken power law with varying power index (red dotted line) and the 
               broken power law with varying power index and oblateness (blue dashed-dotted-
               dotted line) models. 
               The grey areas denote data that have been masked from the fitting to account for 
               the presence of substructure. 
               }
   \label{fig:fits_0.2mag}
   \end{sidewaysfigure*}

   Comparing the parameters resulting from the best fits to the masked $0.2$ mag and $0.4$ mag 
   data, we find that the fits to $0.2$ mag binned data perform better for all the 
   models ($\chi^2_{red}$ ratio of two). Nonetheless, all the measurements for the different 
   structural parameters in the two data sets are compatible with each other within the 
   uncertainties. The best fits for the four models and their residuals for our eight lines 
   of sight are shown in Figures~\ref{fig:fitted_densProfs} and \ref{fig:fits_residuals} for 
   the masked $0.2$ mag binned data. %, and in figures~\ref{fig:fitted_densProfs_0.4} and 
   %\ref{fig:fits_residuals_0.4} for the masked $0.4$ mag binned data. 
   It is clear that the differences between the fitted models along these sight lines are small.

   % Figure: Fitted density profiles 0.2 mag
   %******************************************

   % UNCOMMENT FOR THESIS !!!!!!!!!!!!!!!!!!!!!!!!!!!!!!!!!!!!!!!!!!!!!
   
   % Figure: Fitted density profiles 0.4 mag
   %******************************************
   %\begin{sidewaysfigure*}
   %\centering
   % \subfloat[Fitted density profiles for the $0.4$ mag binned data.]{
   %          \includegraphics[width=12cm]{images/fitteddensityprofile_0.4DMbin_group.eps}
   %          \label{fig:fitted_densProfs_0.4}
   %          }
   % \subfloat[Data-to-model residuals for the $0.4$ mag binned data.]{
   %          \includegraphics[width=12cm]{images/residualsdensitymodels_0.4DMbin_group.eps}
   %          \label{fig:fits_residuals_0.4}
   %          }
   %\caption{Left panels: density profiles in decimal logarithmic scale and the best fit models from 
   %         table~\ref{table:params_0.4mag_nosubstr} (masked $0.4$ binned data). 
   %         Right panels: Residuals between the data and the best fit models from 
   %         panel~\ref{fig:fits_residuals_0.4}. 
   %         The markers and shadowed area are the same as in figure~\ref{fig:fits_0.2mag}.
   %            }
   %\label{fig:fits_0.4mag}
   %\end{sidewaysfigure*}

   Our data are inconclusive regarding triaxiality, but are compatible with either a moderately 
   triaxial halo or with no triaxiality. For the $0.2$ mag data set, 
   the triaxial model fits slightly better than the axisymmetric model and returns $w=0.75\pm0.09$. 
   For the $0.4$ mag data set, however, the axisymmetric model fits slightly better and the 
   triaxial model returns a disk axis ratio compatible with $1$. In both data sets the other 
   best-fitting parameters are practically identical for the two models. This indicates that the 
   cost of the extra parameter is not supported by the $0.4$ mag data. Thus, it is hard to derive 
   a precise value for the disk axis ratio and to conclude if it is truly triaxial, but a 
   weighted average of $w$ and the general analysis show confidently that $w\geq0.8$. 
   
   We increase the complexity of the axisymmetric model by adding two degrees of freedom 
   and considering a change in the power law index $n$ at a specific break distance $R_{break}$ 
   (a broken power law). For this purpose, we use a grid of values to explore all the parameters 
   except the density scale factor $\rho_0$, which we left free to fit (see below for the grid 
   characterization). This model decreases 
   the $\chi^2_{red}$ in both the $0.2$ and the $0.4$ mag binned cases, indicating that 
   our data is better fit by a broken power law than by a simple axisymmetric model or a 
   triaxial model. It turns the single power law index from $n=-4.26\pm0.06$ into a less 
   steep inner index $n_{in}=-2.50\pm0.04$ and a steeper outer index $n_{out}=-4.85\pm0.04$ 
   (measurements here are for weighted averages between the $0.2$ and $0.4$ mag data). It also 
   increases the central value of the polar axis ratio $q$ within the uncertainties, from a 
   weighted $q=0.60\pm0.04$ to a weighted $q=0.63\pm0.02$. Globally, the disk axis ratio seems 
   to be the most stable parameter throughout the different model fits to our data, returning 
   a quite oblate halo. 

   Finally we fix the break distance at the best fit value found by the broken power law model 
   ($R_{break}=19$ kpc and $20$ kpc for the $0.2$ and $0.4$ mag binned data, respectively) and 
   add another parameter to it, allowing not only $n$, but also $q$ to change at the break distance. 
   We find that the best fits to this model return such large error bars for the inner halo that, 
   in practice, it yields unconstrained measurements: $\Delta\rho_0\leq\rho_0$, $\Delta n_{in}$ 
   is 12-18\% of $n_{in}$ and $\Delta q_{in}$ is 40\% of $q_{in}$.
   %We interpret that the more complex broken power law is underconstrained by our data due to a 
   %scarce coverage in the inner halo that, when separated from the outer halo, loses the global 
   %constraing power.

\begin{figure*}
   \centering
    \subfloat[$\chi^2_{red}$ map for the filtered $0.2$ mag binned data set.]{
             \includegraphics[width=8cm]{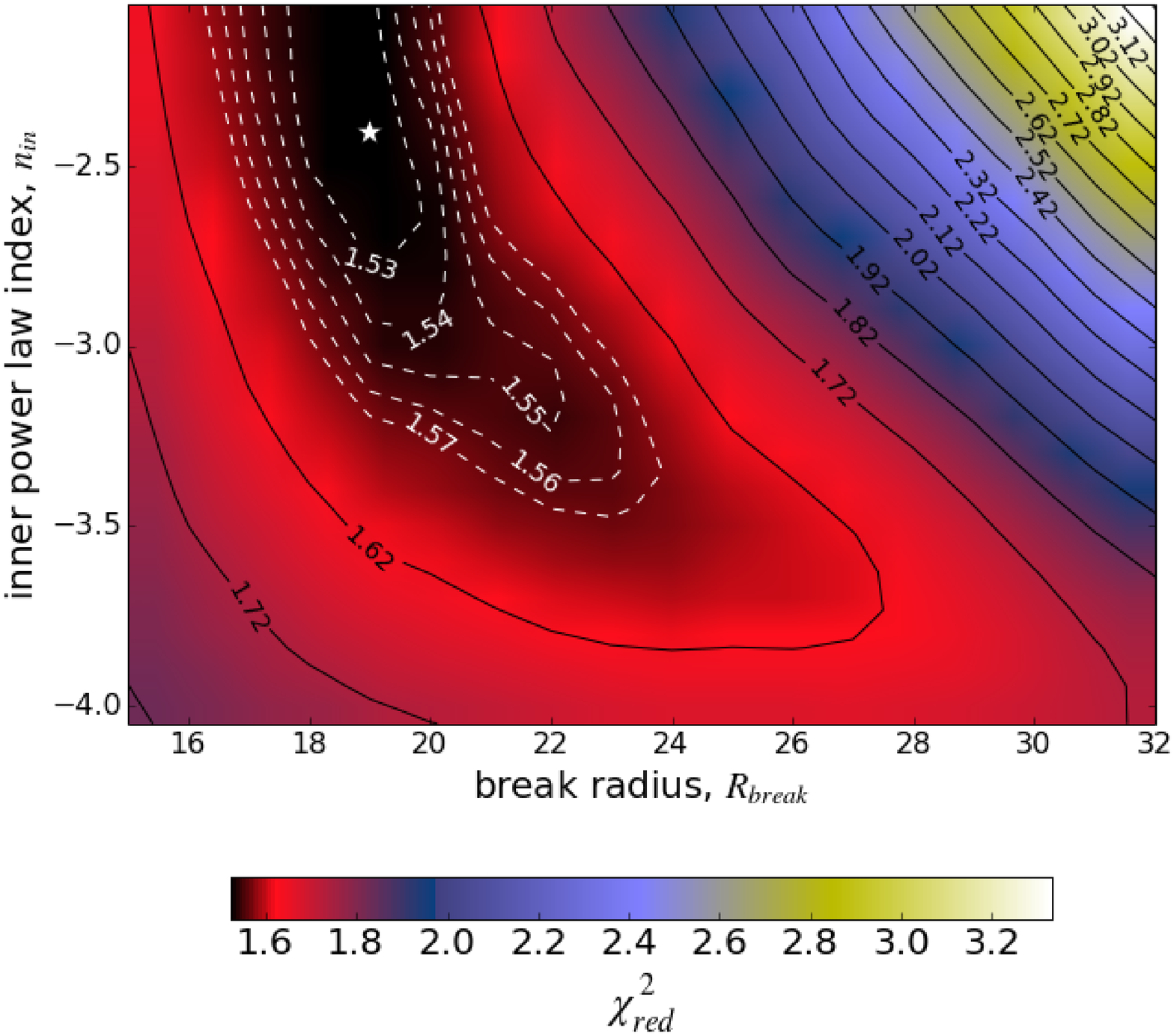}
             \label{fig:chi_map_0.2}
             }
    \subfloat[$\chi^2_{red}$  map for the filtered $0.4$ mag binned data set.]{
             \includegraphics[width=8cm]{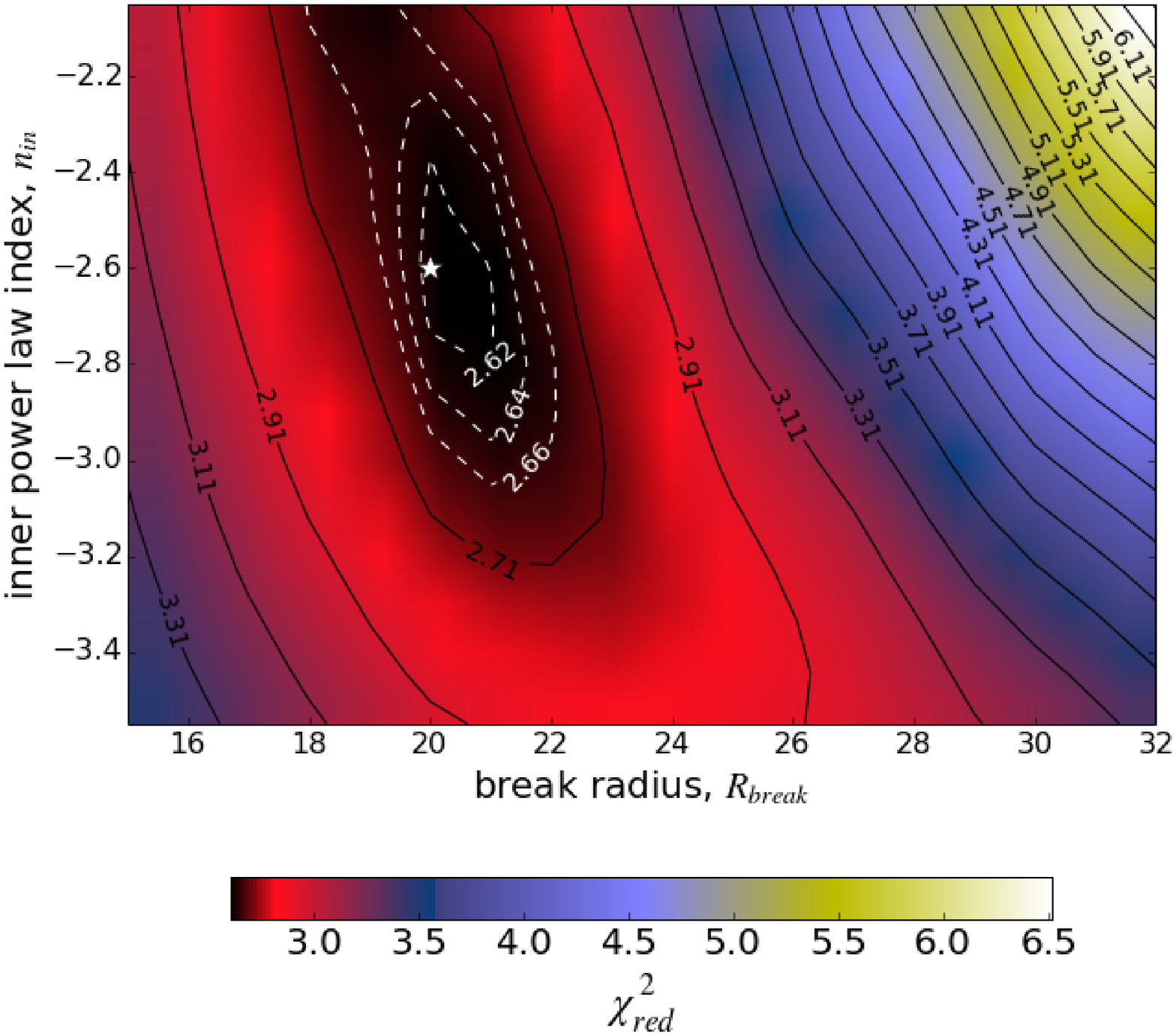}
             \label{fig:chi_map_0.4}
             }
   \caption{$\chi^2_{red}$ isocontours maps for $n_{in}$ and $R_{br}$ from the simple broken 
            power law model. The minimum is indicated with a 
            white star. The black solid isocontours range from $min(\chi^2_{red})+0.1$ to the 
            maximum value, whereas the white dashed isocontours range from $min(\chi^2_{red})+0.01$ 
            to $min(\chi^2_{red})+0.05$. The maps illustrate a degeneracy between both parameters 
            in the best fits.
               }
   \label{fig:chi_maps_simplebroken}
   \end{figure*}

   We explore each model to investigate possible parameter degeneracies, tolerance ranges and 
   potential local minima in our best fits. For this we fix all the parameters in the four models 
   except the density scale factor $\rho_0$, and we run the fits across a grid of parameter values. In 
   particular, the grids are built following $q, w\in[0.1,2.0;\delta=0.05]$, $n\in[-5.0-1.0;\delta=0.1]$, 
   $n_{in}\in[-4.0,-1.0;\delta=0.1]$, $n_{out}\in[-7.0,-3.0;\delta=0.2]$ and $R_{br}\in[15,50;\delta=1]$, 
   where $\delta$ is the incremental step for each parameter. 
   %Through this method we found that the simple broken power law was running into a local minimum 
   %at $R_{br}\approx20$ kpc when fitting the $0.2$ mag binned data set. We solved this by fixing 
   %$R_{br}$ to the best fit value indicated by the grid ($R_{br}=23.0\pm0.5$) and letting the 
   %algorithm find again the best fit parameters for the model. This result and not the one from the 
   %local minimum is the one included in table~\ref{table:params_0.2mag_nosubstr}. As a result of the 
   %grids procedure, we also 
   We find that there is a degeneracy between $R_{br}$ and $n_{in}$ for the simple broken power 
   law model for both binnings (see Figure~\ref{fig:chi_maps_simplebroken}).

   % Figure: chi^2 isocontours maps
   %*********************************

   Finally our measurements for the density scale factor $\rho_0$ ($\rho$ at $R_{GC}=1$ kpc) are 
   the result of large extrapolations and merely serve as normalizations for our fits. For that 
   reason we do not discuss these values in detail. 
   %This is exemplified by the fact that our $\rho_0$ measurements are two to three 
   %orders of magnitude larger than those found by previous studies like that of \citet{sesar11}, 
   %and have large uncertainties, comparable to the values themselves.

%%%%%%%%%%%%%%%%%%%%%%%%%%%%%%%%%%%%%%%%%%%%%%%%%%%%%%%%%%%%%%%%%%%%%%%%%%%%%%%%%%%%%%%%%%%%
%  _________________________________________________________________________________________
% 
%  4. DISCUSSION 
%  _________________________________________________________________________________________
%
%%%%%%%%%%%%%%%%%%%%%%%%%%%%%%%%%%%%%%%%%%%%%%%%%%%%%%%%%%%%%%%%%%%%%%%%%%%%%%%%%%%%%%%%%%%%

\section{Discussion}\label{discussion}

   % ---------------------------------------------------------------------------------------
   % 4.1 ANALYSIS OF THE BEST FITS AND IMPLICATIONS FOR THE GALAXY
   % ---------------------------------------------------------------------------------------

   \subsection{Robustness of the best fit structural parameters}
   \label{subsec:test_bestfits}
   
   In order to determine how the data available to us influences the results from our best fits, 
   we remove the different lines of sight one at a time and repeat the fits. In this way we can 
   determine which are the most critical lines of sight and what is their effect on our results. 
   
   We find that most of them have no significant 
   influence on the best fit parameters of the different halo models. However, starting with the 
   polar axis ratio we find that removing group A significantly increases its value ($q\approx0.7$) 
   and removing groups C or E significantly decreases it ($q\approx0.5$) in both the 
   axisymmetric and triaxial model in the two data sets. Regarding the power law index, again groups 
   A or C have an influence, but group B as well. Removing groups A or B increases $n$ to 
   $\approx-4.1\pm0.1$, whereas removing C decreases it to $n\approx-4.6$. When considering 
   a triaxial halo, we find that groups A, B or C increase the disk axis ratio $w$ by $0.10-0.15$, 
   and that removing groups E or F decreases it to $w\approx0.5$. Additionally, in conditions of 
   triaxiality, the lack of group E reduces $q$ further to $q\approx0.35$. 

   Thus removing group E turns out to be critical for both $q$ and $w$, 
   representing a totally differently looking halo (extremely oblate and quite elliptical in the plane). 
   Group F also has a similar effect on $w$ but not on $q$. 
   The reason why group E has such a strong influence in the determination of a possible 
   triaxiality is that it is by far the closest group to the Galactic anticentre. Other groups 
   also influence the measurements of the different parameters, but have a smaller influence on 
   the general picture we would derive. Overall we see that the lines of sight we use can have a 
   drastic effect on 
   the $w$ results and a significant but moderate effect on $q$ and $n$. This means that 
   a global view of the halo is essential owing to its complex structure.

   % Figure: example of a case of bad fitting
   %*********************************************
   %\begin{sidewaysfigure*}
   %\centering
   % \subfloat[Data-to-model residuals for the triaxial model with free orientation of the x-y axes 
   %           for fixed values of $\rho_0$ and $n$ and constrained parameters.]{
   %          \includegraphics[width=12cm]{images/residuals_rotatingtriax_0.2DMbin.eps}
   %          \label{fig:rotating_residuals}
   %          }
   % \subfloat[Data-to-model residuals for the broken power laws with fixed values of $n_{out}$ 
   %           and $q_{out}$.]{
   %          \includegraphics[width=12cm]{images/residuals_brokenpowerlaws_0.2DMbin.eps}
   %          \label{fig:broken_residuals}
   %          }
   %\caption{\textbf{NEEDS UPDATING WITH AN EXAMPLE OF BAD BITTING}
   %         Residuals between the data (filtered $0.2$ binned data) and the best fit models 
   %         with a smaller number of free parameters. The markers and shadowed area are 
   %         the same as in figure~\ref{fig:fits_0.2mag}. There are no strong differences 
   %         between this residuals and those in \ref{fig:fits_residuals}. 
   %         }
   %\label{fig:fixedparams_0.4mag}
   %\end{sidewaysfigure*}

   % ---------------------------------------------------------------------------------------
   % 4.2 COMPARISON TO PREVIOUS STUDIES
   % ---------------------------------------------------------------------------------------

   \subsection{Comparison to previous studies}\label{subsec:compData}
   
   Previous investigations using near-MSTO stars have explored both the inner and the outer halo 
   out to moderate distances ($30-40$ kpc), and similar regimes have been probed with blue horizontal 
   branch stars and blue struggler stars, MSTO stars or multiple stellar halo tracers. Studies 
   involving RRLyrae stars have reached further out to $50$kpc. Remarkably, the depth of our 
   data allows us to probe further than any previous study (out to $60$ kpc) in several directions, 
   independently of the stellar tracer. 

   In this section we compare our 
   findings regarding the structural parameters of the stellar halo to those of the following 
   results in the literature:
   
   \begin{itemize}
      \item[-] \citet{juric08} use near-MSTO stars from the SDSS-DR3 and DR4 as stellar tracers, 
               and cover the $5\, \mathrm{kpc}<R_{GC}<15$ kpc range. They comprise $5450$ deg$^2$
               in the northern Galactic hemisphere and $1088$ deg$^2$ in the south.
      \item[-] \citet{sesar11} use as well near-MSTO stars from the CFHT Legacy Survey, and 
               explore the  $5\, \mathrm{kpc}<R_{GC}<35$ kpc range. Two of their four fields 
               explore the South Galactic Cap. 
      \item[-] \citet{deason11} use type A blue horizontal branch (BHB) stars and blue 
               stragglers (BS), reaching out to $R_{GC}=40$kpc.
      \item[-] \citet{dejong10} use 
               CMD fitting of SEGUE stellar photometry to probe the total stellar mass density 
               from $R_{GC}=7$ kpc to $R_{GC}=30$ kpc along a "picket fence" of $2.5$ degree 
               wide strips at fixed Galactic longitude spanning a large range of Galactic latitudes. 
               %several SEGUE halo tracers selected by means of CMD fitting. 
               %Their data span from $R_{GC}=7$ kpc to $R_{GC}=30$ kpc at several lines of 
               %sight of narrow galactic longitude and ample galactic latitude (from $b=-10$
               %deg or $b=-40$ deg to $b=50$ deg).
      \item[-] \citet{chen01} use more general MSTO stars from two high latitude regions of 
               SDSS to the North and the South of the Galactic plane ($49\, \mathrm{deg}<|b|<64\, 
               \mathrm{deg}$). They explore the inner halo regime ($R_{GC}\lesssim30$ kpc).
      \item[-] \citet{bell08} use also more general MSTO stars from SDSS-DR5 spanning $5<R_{GC}
               <40$ kpc. 
      \item[-] \citet{faccioli14} use RRLyrae in the $9\, \mathrm{kpc}<R_{GC}<49$ kpc range. 
               Their multiepoch data comes from the Xuyi Schmidt Telescope Photometric Survey 
               (XSTPS) in combination with SDSS colours, and covers $376.75$ deg$^2$ at 
               $RA\approx150$ deg and $Dec\approx27$ deg.
      \item[-] \citet{sesar10densProfile} use RRLyrae stars from SDSS-II in the stripe 82 region.
               Although their data originally spans $5\, \mathrm{kpc}<R_{GC}<110$ kpc, the 
               reanalysis performed by \citet{faccioli14} to derive structural parameters 
               truncates the sample at $49$ kpc.
      \item[-] \citet{watkins09} use as well RRLyrae from SDSS in stripe 82, and the 
               comparative derivation of structural parameters by \citet{faccioli14} also 
               truncates it at $49$ kpc. Stripe 82 is located in the South Galactic Cap. 
   \end{itemize}
   
   The result of this comparison is summarized in Table~\ref{table:compare_params}. %, where our 
   %results are the weighted averages for the best fit over the $0.2$ and $0.4$ mag data sets. 
   We note that the oblateness values for \citet{faccioli14}, 
   \citet{sesar10densProfile} and \citet{watkins09} are not the result of absolute best fits 
   to a set of free parameters, but the best fits to free $R_{br}$, $n_{in}$ and $n_{out}$ 
   with fixed prior values for a quite oblate ($q=0.59^{+0.02}_{-0.03}$) and a moderately 
   oblate halo ($q=0.70\pm0.01$). %They interpreted their similar values of $\chi^2_{red}$ as 
   %indicating the need for larger samples, and as indicating that their $q$ values should 
   %not be taken conclusively. 

   %Most of these works favour an axisymmetric broken power law. Indeed 
   All surveys that reach beyond $R_{GC}=30$ kpc coincide in the need for a break in the 
   power-law index of the halo density. 
   Regarding possible triaxiality, only a few of the studies report constraints on $w$. 
   Those that do, have either reported 'finding unreasonable values' \citep{sesar11} or 
   have obtained limits on triaxiality similar to ours ($w>0.8$, \citet{bell08}). 
   %In relation to a possible triaxiality, most of these works do not qoute any value for the disk 
   %axis ratio $w$, suggesting that either they did not have any reason to test such a model or that 
   %they did but found no indication of it. From those who have reported testing the triaxiality, 
   %\citet{sesar11} found their best fit value to be incompatible with reasonably expected values, 
   %and \citet{bell08} found a lower limit compatible with our findings ($w=0.9\pm0.1 > 0.8$) and 
   %with a non-existing (or very mild) triaxiality.  
   
   On the break radius, there is a general consensus towards $R_{break}\approx27$ kpc. 
   The only exception is that of \citet{bell08}, who find a value very close to our 
   measurement ($\sim20$ kpc). These discrepancies, however, can be explained by the effect 
   of the $R_{break}$-$n_{in}$ degeneracy discussed in section~\ref{subsec:results}.
   
   The inner and outer halo power law indices mostly fall in the $[-2.3,-3.0]$ and 
   $[-3.6,-5.1]$ ranges. Our inner power law index $n_{in}=-2.50\pm0.04$ is consistent with 
   these results, particularly with the lower end. In the case of the outer halo power index 
   ($n_{out}=-4.85\pm0.04$), the comparison is less trivial. First, only 
   \citet{sesar11} and \citet{deason11} 
   have provided measurements for $n_{out}$ based on fits with a free $q$ parameter 
   ($n_{out}=-3.8\pm0.1$ and $-4.6^{+0.2}_{-0.1}$, respectively). 
   Second, only one work with $n_{out}$ measurements \citep{sesar11} uses a stellar 
   tracer similar to ours (the others use A-BHB and BS stars, or RRLyrae stars). 
   Most important, a good constraint on $n_{out}$ requires deep data, and none of these earlier 
   surveys reach as deep as our data set. Our steep outer index, although well in the 
   range of previous measurements, might well indicate a progressive steepening of the halo 
   density, though it would be good to test this with additional sight lines of comparable depth. 
   In any case, it seems safe to conclude that $n_{out}<-4.0$. 
   %And third, our 
   %data are the one probing the halo the furthest (twice as far as the next furthest case), with 
   %possible implications for the stability and the average of $n_{out}$ (as suggested by the 
   %distance threshold test and figure~\ref{fig:distThr}). For all these reasons we feel inclined 
   %to suggest deeper surveys --if possible, with similar stellar tracers-- and model fits with 
   %a freely-varying oblateness before settling on any prefered value for $n_{out}$. Nonetheless, 
   %it seems safe to derive from comparing all the different works that $n_{out}<-3.5$ and 
   %probably $<-4.0$.
   
   The best fit values for the polar axis ratio or oblateness $q$ range from $0.5$ 
   to $0.9$, with most of the measurements concentrated within $(0.55,0.70)$. The values of $q$ 
   do not seem to depend on whether a break was detected or not, nor on the limiting 
   distance of the survey or on the stellar tracer. The discrepancies can thus be attributed either 
   to methodological differences or to differences in the spatial coverage of the data samples. 
   However, it is difficult to determine the actual cause. Our results fit well within the most 
   constricted range, and the simple broken power law measure (which is the one with the best 
   $\chi^2_{red}$ and $q=0.63\pm0.04$) falls near the upper end, being in good agreement with 
   \citet{juric08}, \citet{sesar11} and close to \citet{deason11} and \citet{chen01}.  
   %In keeping with this line of thought, it is worth cautioning that typical MSTO and near-MSTO 
   %star selections can be contaminated by quasars and white dwarf-M dwarf pairs if further 
   %filtering is not applied. Due to the specific absolute brightness and spatial distribution 
   %of these two types of sources, their effect on the stellar distribution profiles 
   %might not be negligible.
   
   Finally it is noteworthy that the choice of stellar tracer across the different works 
   does not seem to cause any significant bias on the best fit parameters.

   % Table 5: comparison of parameters to previous works
   %*******************************************************
   \begin{sidewaystable*}
   \caption{Comparison between the best fit structural parameters found in this work (weighted 
            averages for the parameters of the $0.2$ and $0.4$ mag data sets) and 
            those reported by other groups in previous works. The different works have been 
            labelled as follows: J08 \citep{juric08}, S11 \citep{sesar11}, D11 \citep{deason11}, 
            dJ10 \citep{dejong10}, Ch01 \citep{chen01}, B08 \citep{bell08}, F14 \citep{faccioli14}, 
            and S10 \citep{sesar10densProfile} and W09 \citep{watkins09} as reanalysed in F14. 
            %In general the models measure a single 
            %value for the oblateness both for the inner and outer halo, unless otherwise stated. 
            The fitted models in F14, S10 and W09 have fixed oblateness and test two different 
            values motivated by the previous findings in S11 and D11.
            }                      
   \label{table:compare_params}                    % is used to refer this table in the text
   \centering                                                     % used for centering table
   %\begin{tabular}{l l c c c c c c c c c}                    % centered columns (4 columns)
   \begin{tabular}{l l c c c c c c c c c}                     % centered columns (4 columns)
   \hline%\hline                                           % inserts double horizontal lines
                                                                             % table heading 
    Work & stellar tracer & dist. range (kpc) & $\chi_{red}^2$ & $R_{br}\, \mathrm{(kpc)}$ 
    & $n$ & $n_{in}$ & $n_{out}$ & $q$ & $w$ \\
   \hline                                                  % inserts single horizontal line
                                                                        % body of the table
   this work-axisym. & near-MSTO & $[10,60]$ & 1.9 %& $(6.0\pm0.4)\cdot 10^{-4}$ & -- 
                     & -- 
                     & $-4.28\pm0.06$ & -- & -- 
                     & $0.61\pm0.04$ &  --  \\ 
   this work-triax. & near-MSTO & $[10,60]$ & 1.9 %& $(6.0\pm0.4)\cdot 10^{-4}$ & -- 
                     & -- 
                     & $-4.26\pm0.06$ & -- & -- 
                     & $0.60\pm0.04$ &  $0.81\pm0.07$  \\ 
   this work-broken & near-MSTO & $[10,60]$ & 1.5 %& $(6.0\pm0.4)\cdot 10^{-4}$ & -- 
                     & $19.5\pm0.4$ 
                     & -- & $-2.50\pm0.04$ & $-4.85\pm0.04$ 
                     & $0.63\pm0.02$ &  --  \\ 
   \hline    
   J08 & near-MSTO & $[5,15]$ & $[2,3]$ %& $x\cdot 10^{-6}$ 
       & -- 
       & -- & $-2.8\pm0.3$ & -- 
       & $0.65\pm0.15$ & -- \\ 
   S11 & near-MSTO & $[5,35]$ & $3.9$ %& $[1.40,1.70]\cdot 10^{-6}$ 
       & $27.8\pm0.8$ 
       & -- & $-2.62\pm0.04$ & $-3.8\pm0.1$ 
       & $0.70\pm0.02$ &  \tiny{excluded}  \\ 
   D11 & A-BHB, -BS & $[-,40]$ & -- %& $x\cdot 10^{-6}$ 
        & $27.1\pm1$ 
        & -- & $-2.3\pm0.1$ & $-4.6^{+0.2}_{-0.1}$ 
        & $0.59^{+0.02}_{-0.03}$ &  -- \\ 
   dJ10 & multiple & $[7,30]$ & $[3.9,4.2]$ %& $x\cdot 10^{-6}$ 
        & -- 
        & $-2.75\pm0.07$ & -- & -- 
        & $0.88\pm0.03$ & -- \\ 
   Ch01 & MSTO & $[-,30]$ & -- %& $x\cdot 10^{-6}$ 
        & -- 
        & $-2.5\pm0.3$ & -- & -- 
        & $0.55\pm0.06$ &  -- \\ 
   B08 & MSTO & $[5,40]$ & $2.2$ %& $x\cdot 10^{-6}$ 
       & $\sim20$ 
       & $-3\pm1$ & -- & -- 
       & $[0.5,0.8]$ &  $\geq0.8$  \\ 
   F14 & RRLyrae & $[9,49]$ & $0.8$ %& $n_0=(0.21\pm0.14)kpc^{-3}$ %$x\cdot 10^{-6}$ 
       & $28.5\pm5.6$ 
       & -- & $-2.8\pm0.4$ & $-4.4\pm0.7$ 
       & $q_{fix}=0.70\pm0.01$ &  --  \\ 
    "  & RRLyrae & $[9,49]$ & $1.04$ %& $n_0=(0.21\pm0.14)kpc^{-3}$ %$x\cdot 10^{-6}$ 
       & $26.5\pm8.9$ 
       & -- & $-2.7\pm0.6$ & $-3.6\pm0.4$ 
       & $q_{fix}=0.59^{+0.02}_{-0.03}$ &  --  \\ 
   S10 & RRLyrae & $[9,49]$ & $1.1$ %& $x\cdot 10^{-6}$ 
       & $34.6\pm2.8$ 
       & -- & $-2.8\pm0.2$ & $-5.8\pm0.9$ 
       & $q_{fix}=0.70\pm0.01$ &  --  \\ 
    "  & RRLyrae & $[9,49]$ & $1.52$ %& $x\cdot 10^{-6}$ 
       & $26.2\pm7.4$ 
       & -- & $-3.0\pm0.3$ & $-3.8\pm0.3$ 
       & $q_{fix}=0.59^{+0.02}_{-0.03}$ &  --  \\ 
   W09 & RRLyrae & $[9,49]$  & $1.1$ 
       & $27.6\pm3.3$ 
       & -- & $-2.5\pm0.3$ & $-4.3\pm0.4$ 
       & $q_{fix}=0.70\pm0.01$ &  --  \\ 
    "  & RRLyrae & $[9,49]$ & $0.69$  
       & $26.9\pm3.1$ 
       & -- & $-2.1\pm0.3$ & $-4.0\pm0.3$ 
       & $q_{fix}=0.59^{+0.02}_{-0.03}$ &  --  \\ 
   \hline                                                             %inserts single line
   \end{tabular}
   \end{sidewaystable*}

   % ---------------------------------------------------------------------------------------
   % 4.3 OVERDENSITIES AND SUBSTRUCTURE
   % ---------------------------------------------------------------------------------------

   \subsection{Detection of overdensities and identification}\label{subsec:compModels}
   
   We analyse the data-to-models residuals for the different lines of sight in 
   Figure~\ref{fig:fits_residuals} in search for overdensities. 
   We find that, in general, all the lines of  sight present regions with data-to-models 
   deviations of a maximum factor of two. 
   Additionally, certain lines of sight --C,D, G, and H-- present more significant deviations 
   spanning from a few kiloparsecs to tens of kiloparsecs in distance. We discuss these 
   overdensities in greater detail below, and we also discuss expected overdensities that 
   show no signature in our data. 

   % Figure: Equatorial map with streams
   %***************************************
   \begin{figure*}
   \centering
   \includegraphics[width=\textwidth]{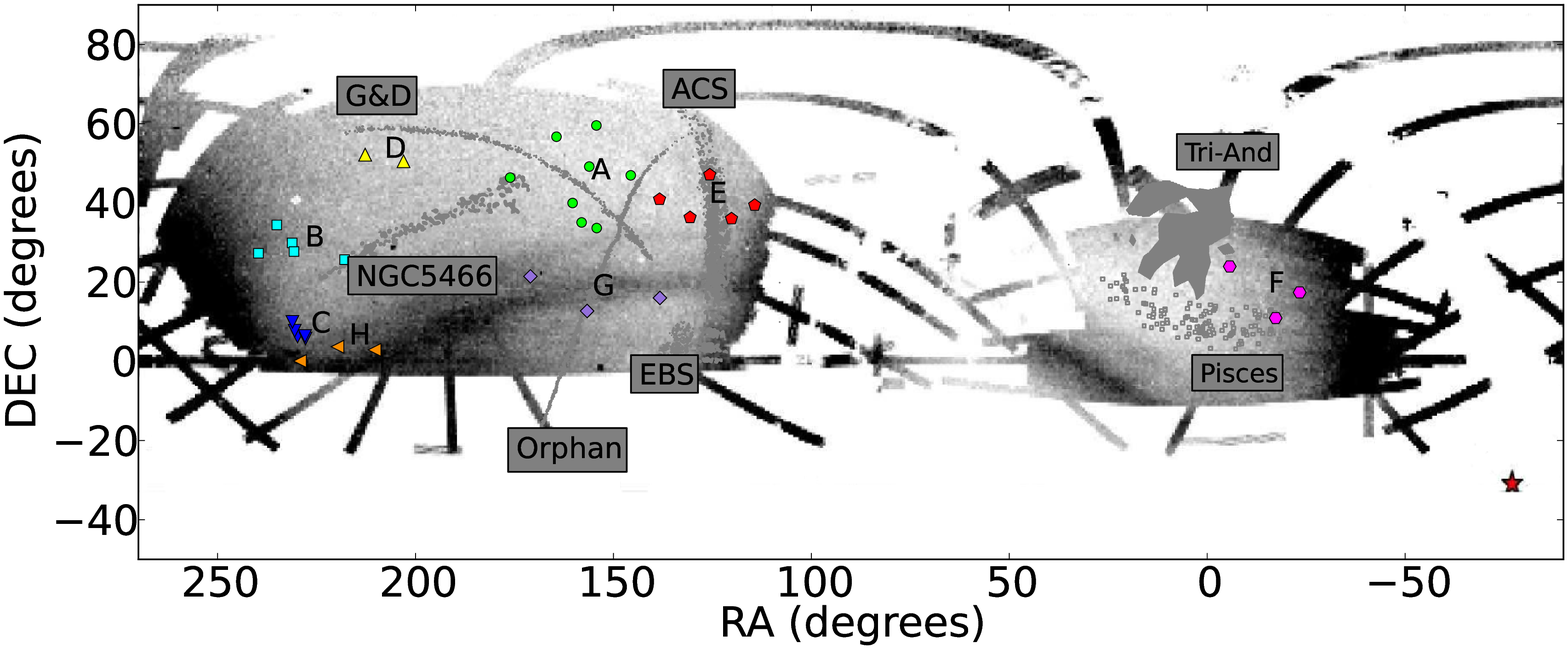}
      \caption{Equatorial map showing the position of all the fields used in this work and 
               the closest cold stellar overdensities to them. These overdensities are 
               used for comparison and discussion of the stellar density profile data-to-model 
               residuals throughout section~\ref{subsec:compModels}. The labels in the figure 
               correspond to the Anticentre Structure (ACS), the Eastern Band Structure (EBS), 
               the NGC5466 stream, the Grillmair \& Dionatos stream (G\&D), the Orphan stream, 
               the Triangulum-Andromeda overdensity (Tri-And) and the Pisces overdensity. 
               %The pointings englobed with a black ellipse are those contributing to the 
               %overdensity in group A. 
               The background image is the SDSS-DR8 map from \citet{kopos12}, which shows the 
               footprint of the Sagittarius stream. The Monoceros ring also appears partially 
               in this background image, as a dark region overlapping the western part of the 
               Galactic disk in the anticentre region, eastwards of the ACS. 
               }
         \label{fig:mapStreams}
   \end{figure*}

   The most prominent overdensities in the data-to-model residuals correspond to the northern 
   wrap of the Sagittarius (Sgr) stream. This stream overlaps in projection with groups G and H (see 
   Figure~\ref{fig:mapStreams}). 
   For group G, the residuals indicate overdensities in the distance range where we expect to 
   find both the Sgr and the Orphan stream ($20<D_{hC}\lesssim40$ kpc or $25<D_{GC}\lesssim44$ 
   kpc, \citet{piladiez14}). The overdensities indeed peak between $R_{GC}=25$ kpc and $45$ kpc, 
   reaching $\rho/\rho_M=7\pm2$, and drop sharply afterwards. Group H probes the Sgr 
   stream closer to the Galactic centre but also for larger distances than group G. Based both on 
   extensive data (summarized in \citet{piladiez14}) and in models (\citet{law10} and 
   \citet{penarrub10}), we expect this stream to span the $20<D_{hC}<60$ kpc or $16<R_{GC}<55$ 
   kpc range at these coordinates. This expectation is met all along: they steadily increase 
   from $R_{GC}\approx15$ kpc, depart from $\rho/\rho_M=3\pm1$ at $R_{GC}=30$ kpc, reach 
   $\rho/\rho_M=6\pm2$ at $R_{GC}=40$ kpc and peak at $R_{GC}=45$ kpc with $max(\rho/\rho_M)=(12,15)\pm2$.  
   However, they do not decrease near $R_{GC}=55$ kpc but seem to stay stable 
   with a significant $\rho/\rho_M>7\pm2$). This suggests a 
   thicker branch than predicted by the models, but in agreement with previous RRLyrae 
   measurements (\citet{ibata01carbonstars}, \citet{totten98carbonstars} and \citet{dp01} as 
   summarized in Figure~17 of \citet{maj03}).

   Two more modest overdensities that do not appear in the literature seem to be present in 
   groups C and D. In group C, a weak but consistent overdensity spans a distance range of 
   $R_{GC}\approx35$ kpc to $R_{GC}\approx60$ kpc. In group D, a sharp bump extends over a 
   few kiloparsecs around $R_{GC}\leq20$ kpc.

   We have looked for other known overdensities that position-match our lines of sight (see 
   Figure~\ref{fig:mapStreams}), 
   but found no indication of them in the residuals. The first one corresponds to the 
   tidal tails of the NGC5466 globular cluster \citep{belokurovNGC5466}, which overlap 
   with one field in group A and another one in group B (A1361 centred at 
   $(RA,Dec)=(176.09, 46.39)$ and A1927 at $(RA,Dec)=(217.92, 25.67)$). 
   This is a very weak cold substructure located at $R_{GC}\approx=16$ kpc and extending for 
   $45\deg$ with an average width of $1.4\deg$ \citep{grillmairjohnson06_ngc5466}. As such, 
   it is not surprising to find no signature in the density profiles.

   The second one is the ensemble of three known overdensities in the direction of group E: 
   the Anti Center Stream ($R_{GC}=18\pm2$ kpc, \citet{rochapinto03} and \citet{li12}), 
   the Monoceros ring ($R_{GC}\approx18$ kpc, \citet{li12}) and the Eastern Band Structure 
   ($R_{GC}=20\pm2$ kpc, \citet{li12}). These substructures are masked from our fits 
   and residuals when we impose $|z|>10$ kpc to avoid the influence of the thick disk, 
   and therefore, they cannot be detected. 

   The Triangulum-Andromeda overdensity (\citep{martin07}) falls close to one of the fields 
   in group F. Despite this proximity, the residuals show no evidence for an overdensity at the 
   expected distance of $R_{GC}\approx30$ kpc, 
   indicating that the overdensity does not extend further in this direction.

%%%%%%%%%%%%%%%%%%%%%%%%%%%%%%%%%%%%%%%%%%%%%%%%%%%%%%%%%%%%%%%%%%%%%%%%%%%%%%%%%%%%%%%%%%%%
%  _________________________________________________________________________________________
% 
%  5. CONCLUSIONS
%  _________________________________________________________________________________________
%
%%%%%%%%%%%%%%%%%%%%%%%%%%%%%%%%%%%%%%%%%%%%%%%%%%%%%%%%%%%%%%%%%%%%%%%%%%%%%%%%%%%%%%%%%%%%

\section{Conclusions}\label{concl}
   
   In this paper we have used wide-field images from the CFHT and the INT telescopes in 
   eight broad lines of sight spread across the sky to produce deep photometric catalogues 
   of halo near main sequence 
   turnoff (near-MSTO) stars. Our images have been corrected for PSF inhomogeneities, 
   resulting in catalogues with fixed-aperture colour measurements and improved star-galaxy 
   separation. Thanks to the depth and quality of our data, we reach stellar completeness 
   limits ranging from $22.7$ mag to $24.2$ mag in the $r$ band, which translate 
   into a $60$ kpc distance limit for near-MSTO stars.
              
   We calculate galactocentric distances for the stars based on the photometric 
   parallax method by \citet{ivezic08} and the metallicity estimator by \citet{bond10}. 
   We bin them by distance modulus, and calculate 
   the stellar number density distribution along the eight different lines of sight.

   In selecting the halo near-MSTO stars, we have used additional constraints than the standard 
   $0.2<g-r<0.3$ and $g,r,i>17$ cuts in order to obtain a cleaner sample. Particularly, 
   by applying additional cuts based on g-i colour, absolute magnitude and metallicity, 
   we get a sample of mainly F stars significantly decontaminated from quasars and 
   white dwarf-M dwarf pairs. 
              
   We fit several galactic halo models of the stellar distribution to our eight lines of sight, 
   and explore the structural parameters resulting from the best fits, as well as the 
   influence of substructure in those parameters. %We find that our data is most 
   %effective at probing the outer halo, and so we are able to constrain 
   %its structural parameters better than those of the inner halo. 
   We find that the halo is best represented by a broken power law with index $n_{in}=-2.50\pm0.04$ 
   in the inner halo ($R<R_{break}=19.5\pm0.04$) and $n_{out}=-4.85\pm0.04$ in the outer halo. 
   Our data cannot constrain whether a change in the polar axis ratio also accompanies the 
   break in the halo. The best fit values for the polar axes ratio indicate a quite oblate halo: 
   $q=0.63\pm0.02$. The simpler (non-broken) triaxial power law models favour a practically 
   axisymmetric halo, with $w\geq0.81\pm0.07$ and the rest of parameters equal to those of the 
   axisymmetric one. %Finally, we find that the break takes place at $R_{break}=22.8\pm0.5$, but also 
   %that there exists a degeneracy between $R_{break}$ and $n_{in}$ that spans several kiloparsecs. 
   
   We find that fitting models to data that contains substantial substructure can bias significantly 
   the perception of triaxiality, decreasing the disk axis ratio $w$ by $15\%$. We also find that 
   different distance modulus bin sizes and the inclusion or exclusion of particular lines of sight 
   can moderately influence our measurements of some structural parameters. This calls for carefully 
   crafted analysis and tailored tests in any future studies. When compared to previous works, 
   the choice of stellar tracer seems to have no significant influence on the values of the structural 
   parameters, at least for these distance ranges.
   
   Comparing our density profiles to the smooth model fits, we recover the presence of the 
   Sagittarius stream in groups G and H. The Sagittarius stream in the direction of group H seems 
   to extend further out from the Galactic centre than the models have so far predicted, and 
   confirms previous RRLyrae detections associated with the stream at such distances 
   (\citet{ibata01carbonstars}, \citet{totten98carbonstars} and \citet{dp01}). 
   We also find evidence of more modest substructures extending over a long range of distances 
   in group C ($35\leq R_{GC}\leq60$ kpc) and quite concentrated in distance in group D 
   ($R_{GC}\approx20$ kpc). 
   
   Our pencil beam survey has demonstrated that even a relatively small numbers of narrow fields 
   of view, provided they are sampled sufficiently deep and with an abundant tracer, can place 
   competitive limits on the global density profile and shape of the Galactic halo. The advent 
   of similarly deep, wide-area surveys -like KiDS, VIKING and LSST- 
   therefore promises to enhance substantially our understanding of the halo.

%%%%%%%%%%%%%%%%%%%%%%%%%%%%%%%%%%%%%%%%%%%%%%%%%%%%%%%%%%%%%%%%%%%%%%%%%%%%%%%%%%%%%%%%%%%%
%  _________________________________________________________________________________________
% 
%  A. AKNOWLEDGEMENTS
%  _________________________________________________________________________________________
%
%%%%%%%%%%%%%%%%%%%%%%%%%%%%%%%%%%%%%%%%%%%%%%%%%%%%%%%%%%%%%%%%%%%%%%%%%%%%%%%%%%%%%%%%%%%%

\begin{acknowledgements}
      
   %Anonymous referee. 
   This work is based on observations made with the Isaac Newton Telescope 
   through program IDs I10AN006, I10AP005, I10BN003, I10BP005, I11AN009, 
   I11AP013. The Isaac Newton Telescope is operated on the island of La Palma 
   by the Isaac Newton Group in the Spanish Observatorio del Roque de los 
   Muchachos of the Instituto de Astrofísica de Canarias. 
   It is also based on observations obtained with MegaPrime/MegaCam, a joint 
   project of CFHT and CEA/IRFU, at the Canada-France-Hawaii Telescope (CFHT). 
   The CFHT is operated by the National Research Council (NRC) of Canada, the 
   Institut National des Science de l'Univers of the Centre National de la 
   Recherche Scientifique (CNRS) of France, and the University of Hawaii. \\
   
   Most of the data processing and analysis in this work has been carried 
   out using Python and, in particular, the open source modules Scipy, 
   Numpy, AstroAsciiData and Matplotlib. We have also used the Stilts 
   program for table manipulation. \\
   
   We acknowledge the anonymous referee for the constructive suggestions and insightful 
   discussion. B.P.D. is supported by NOVA, the Dutch Research School of Astronomy.
   H.H. and R.vdB. acknowledge support from the Netherlands Organisation for 
   Scientific Research (NWO) grant number 639.042.814.
      
\end{acknowledgements}

%%%%%%%%%%%%%%%%%%%%%%%%%%%%%%%%%%%%%%%%%%%%%%%%%%%%%%%%%%%%%%%%%%%%%%%%%%%%%%%%%%%%%%%%%%%%
%  _________________________________________________________________________________________
% 
%  B. BIBLIOGRAPHY
%  _________________________________________________________________________________________
%
%%%%%%%%%%%%%%%%%%%%%%%%%%%%%%%%%%%%%%%%%%%%%%%%%%%%%%%%%%%%%%%%%%%%%%%%%%%%%%%%%%%%%%%%%%%%

\bibliographystyle{aa}  % style aa.bst
\bibliography{references} % the file containing the references: references.bib

\begin{thebibliography}{40}
\expandafter\ifx\csname natexlab\endcsname\relax\def\natexlab#1{#1}\fi

\bibitem[{{Ahn} {et~al.}(2014){Ahn}, {Alexandroff}, {Allende Prieto}, {Anders},
  {Anderson}, {Anderton}, {Andrews}, {Aubourg}, {Bailey}, {Bastien}, \&
  et~al.}]{ahn14}
{Ahn}, C.~P., {Alexandroff}, R., {Allende Prieto}, C., {et~al.} 2014, \apjs,
  211, 17

\bibitem[{{Bell} {et~al.}(2008){Bell}, {Zucker}, {Belokurov}, {Sharma},
  {Johnston}, {Bullock}, {Hogg}, {Jahnke}, {de Jong}, {Beers}, {Evans},
  {Grebel}, {Ivezi{\'c}}, {Koposov}, {Rix}, {Schneider}, {Steinmetz}, \&
  {Zolotov}}]{bell08}
{Bell}, E.~F., {Zucker}, D.~B., {Belokurov}, V., {et~al.} 2008, \apj, 680, 295

\bibitem[{{Belokurov} {et~al.}(2006{\natexlab{a}}){Belokurov}, {Evans},
  {Irwin}, {Hewett}, \& {Wilkinson}}]{belokurovNGC5466}
{Belokurov}, V., {Evans}, N.~W., {Irwin}, M.~J., {Hewett}, P.~C., \&
  {Wilkinson}, M.~I. 2006{\natexlab{a}}, \apjl, 637, L29

\bibitem[{{Belokurov} {et~al.}(2007){Belokurov}, {Evans}, {Irwin},
  {Lynden-Bell}, {Yanny}, {Vidrih}, {Gilmore}, {Seabroke}, {Zucker},
  {Wilkinson}, {Hewett}, {Bramich}, {Fellhauer}, {Newberg}, {Wyse}, {Beers},
  {Bell}, {Barentine}, {Brinkmann}, {Cole}, {Pan}, \& {York}}]{belok07orphan}
{Belokurov}, V., {Evans}, N.~W., {Irwin}, M.~J., {et~al.} 2007, \apj, 658, 337

\bibitem[{{Belokurov} {et~al.}(2006{\natexlab{b}}){Belokurov}, {Zucker},
  {Evans}, {Gilmore}, {Vidrih}, {Bramich}, {Newberg}, {Wyse}, {Irwin},
  {Fellhauer}, {Hewett}, {Walton}, {Wilkinson}, {Cole}, {Yanny}, {Rockosi},
  {Beers}, {Bell}, {Brinkmann}, {Ivezi{\'c}}, \& {Lupton}}]{belok06}
{Belokurov}, V., {Zucker}, D.~B., {Evans}, N.~W., {et~al.} 2006{\natexlab{b}},
  \apjl, 642, L137

\bibitem[{{Bertin} \& {Arnouts}(1996)}]{bertinSExtractor}
{Bertin}, E. \& {Arnouts}, S. 1996, \aaps, 117, 393

\bibitem[{{Bildfell} {et~al.}(2012){Bildfell}, {Hoekstra}, {Babul}, {Sand},
  {Graham}, {Willis}, {Urquhart}, {Mahdavi}, {Pritchet}, {Zaritsky}, {Franse},
  \& {Langelaan}}]{bildfell12combined}
{Bildfell}, C., {Hoekstra}, H., {Babul}, A., {et~al.} 2012, \mnras, 425, 204

\bibitem[{{Bond} {et~al.}(2010){Bond}, {Ivezi{\'c}}, {Sesar}, {Juri{\'c}},
  {Munn}, {Kowalski}, {Loebman}, {Ro{\v s}kar}, {Beers}, {Dalcanton},
  {Rockosi}, {Yanny}, {Newberg}, {Allende Prieto}, {Wilhelm}, {Lee},
  {Sivarani}, {Majewski}, {Norris}, {Bailer-Jones}, {Re Fiorentin}, {Schlegel},
  {Uomoto}, {Lupton}, {Knapp}, {Gunn}, {Covey}, {Allyn Smith}, {Miknaitis},
  {Doi}, {Tanaka}, {Fukugita}, {Kent}, {Finkbeiner}, {Quinn}, {Hawley},
  {Anderson}, {Kiuchi}, {Chen}, {Bushong}, {Sohi}, {Haggard}, {Kimball},
  {McGurk}, {Barentine}, {Brewington}, {Harvanek}, {Kleinman}, {Krzesinski},
  {Long}, {Nitta}, {Snedden}, {Lee}, {Pier}, {Harris}, {Brinkmann}, \&
  {Schneider}}]{bond10}
{Bond}, N.~A., {Ivezi{\'c}}, {\v Z}., {Sesar}, B., {et~al.} 2010, \apj, 716, 1

\bibitem[{{Chen} {et~al.}(2001){Chen}, {Stoughton}, {Smith}, {Uomoto}, {Pier},
  {Yanny}, {Ivezi{\'c}}, {York}, {Anderson}, {Annis}, {Brinkmann}, {Csabai},
  {Fukugita}, {Hindsley}, {Lupton}, {Munn}, \& {SDSS Collaboration}}]{chen01}
{Chen}, B., {Stoughton}, C., {Smith}, J.~A., {et~al.} 2001, \apj, 553, 184

\bibitem[{{Covey} {et~al.}(2007){Covey}, {Ivezi{\'c}}, {Schlegel},
  {Finkbeiner}, {Padmanabhan}, {Lupton}, {Ag{\"u}eros}, {Bochanski}, {Hawley},
  {West}, {Seth}, {Kimball}, {Gogarten}, {Claire}, {Haggard}, {Kaib},
  {Schneider}, \& {Sesar}}]{covey07}
{Covey}, K.~R., {Ivezi{\'c}}, {\v Z}., {Schlegel}, D., {et~al.} 2007, \aj, 134,
  2398

\bibitem[{{de Jong} {et~al.}(2010){de Jong}, {Yanny}, {Rix}, {Dolphin},
  {Martin}, \& {Beers}}]{dejong10}
{de Jong}, J.~T.~A., {Yanny}, B., {Rix}, H.-W., {et~al.} 2010, \apj, 714, 663

\bibitem[{{Deason} {et~al.}(2011){Deason}, {Belokurov}, \& {Evans}}]{deason11}
{Deason}, A.~J., {Belokurov}, V., \& {Evans}, N.~W. 2011, \mnras, 416, 2903

\bibitem[{{Dohm-Palmer} {et~al.}(2001){Dohm-Palmer}, {Helmi}, {Morrison},
  {Mateo}, {Olszewski}, {Harding}, {Freeman}, {Norris}, \& {Shectman}}]{dp01}
{Dohm-Palmer}, R.~C., {Helmi}, A., {Morrison}, H., {et~al.} 2001, \apjl, 555,
  L37

\bibitem[{{Erben} {et~al.}(2009){Erben}, {Hildebrandt}, {Lerchster}, {Hudelot},
  {Benjamin}, {van Waerbeke}, {Schrabback}, {Brimioulle}, {Cordes}, {Dietrich},
  {Holhjem}, {Schirmer}, \& {Schneider}}]{erben09}
{Erben}, T., {Hildebrandt}, H., {Lerchster}, M., {et~al.} 2009, \aap, 493, 1197

\bibitem[{{Faccioli} {et~al.}(2014){Faccioli}, {Smith}, {Yuan}, {Zhang}, {Liu},
  {Zhao}, \& {Yao}}]{faccioli14}
{Faccioli}, L., {Smith}, M.~C., {Yuan}, H.-B., {et~al.} 2014, \apj, 788, 105

\bibitem[{{Grillmair}(2006)}]{grillmair06anticenter}
{Grillmair}, C.~J. 2006, \apjl, 651, L29

\bibitem[{{Grillmair} \& {Johnson}(2006)}]{grillmairjohnson06_ngc5466}
{Grillmair}, C.~J. \& {Johnson}, R. 2006, \apjl, 639, L17

\bibitem[{{Hildebrandt} {et~al.}(2009){Hildebrandt}, {Pielorz}, {Erben}, {van
  Waerbeke}, {Simon}, \& {Capak}}]{hildebrandt09}
{Hildebrandt}, H., {Pielorz}, J., {Erben}, T., {et~al.} 2009, \aap, 498, 725

\bibitem[{{Hoekstra} {et~al.}(2012){Hoekstra}, {Mahdavi}, {Babul}, \&
  {Bildfell}}]{hoekstra12cccp}
{Hoekstra}, H., {Mahdavi}, A., {Babul}, A., \& {Bildfell}, C. 2012, \mnras,
  427, 1298

\bibitem[{{Ibata} {et~al.}(2001){Ibata}, {Lewis}, {Irwin}, {Totten}, \&
  {Quinn}}]{ibata01carbonstars}
{Ibata}, R., {Lewis}, G.~F., {Irwin}, M., {Totten}, E., \& {Quinn}, T. 2001,
  \apj, 551, 294

\bibitem[{{Ivezi{\'c}} {et~al.}(2008){Ivezi{\'c}}, {Sesar}, {Juri{\'c}},
  {Bond}, {Dalcanton}, {Rockosi}, {Yanny}, {Newberg}, {Beers}, {Allende
  Prieto}, {Wilhelm}, {Lee}, {Sivarani}, {Norris}, {Bailer-Jones}, {Re
  Fiorentin}, {Schlegel}, {Uomoto}, {Lupton}, {Knapp}, {Gunn}, {Covey},
  {Smith}, {Miknaitis}, {Doi}, {Tanaka}, {Fukugita}, {Kent}, {Finkbeiner},
  {Munn}, {Pier}, {Quinn}, {Hawley}, {Anderson}, {Kiuchi}, {Chen}, {Bushong},
  {Sohi}, {Haggard}, {Kimball}, {Barentine}, {Brewington}, {Harvanek},
  {Kleinman}, {Krzesinski}, {Long}, {Nitta}, {Snedden}, {Lee}, {Harris},
  {Brinkmann}, {Schneider}, \& {York}}]{ivezic08}
{Ivezi{\'c}}, {\v Z}., {Sesar}, B., {Juri{\'c}}, M., {et~al.} 2008, \apj, 684,
  287

\bibitem[{{Juri{\'c}} {et~al.}(2008){Juri{\'c}}, {Ivezi{\'c}}, {Brooks},
  {Lupton}, {Schlegel}, {Finkbeiner}, {Padmanabhan}, {Bond}, {Sesar},
  {Rockosi}, {Knapp}, {Gunn}, {Sumi}, {Schneider}, {Barentine}, {Brewington},
  {Brinkmann}, {Fukugita}, {Harvanek}, {Kleinman}, {Krzesinski}, {Long},
  {Neilsen}, {Nitta}, {Snedden}, \& {York}}]{juric08}
{Juri{\'c}}, M., {Ivezi{\'c}}, {\v Z}., {Brooks}, A., {et~al.} 2008, \apj, 673,
  864

\bibitem[{{Koposov} {et~al.}(2012){Koposov}, {Belokurov}, {Evans}, {Gilmore},
  {Gieles}, {Irwin}, {Lewis}, {Niederste-Ostholt}, {Pe{\~n}arrubia}, {Smith},
  {Bizyaev}, {Malanushenko}, {Malanushenko}, {Schneider}, \& {Wyse}}]{kopos12}
{Koposov}, S.~E., {Belokurov}, V., {Evans}, N.~W., {et~al.} 2012, \apj, 750, 80

\bibitem[{{Law} \& {Majewski}(2010)}]{law10}
{Law}, D.~R. \& {Majewski}, S.~R. 2010, \apj, 714, 229

\bibitem[{{Li} {et~al.}(2012){Li}, {Newberg}, {Carlin}, {Deng}, {Newby},
  {Willett}, {Xu}, \& {Luo}}]{li12}
{Li}, J., {Newberg}, H.~J., {Carlin}, J.~L., {et~al.} 2012, \apj, 757, 151

\bibitem[{{Majewski} {et~al.}(2003){Majewski}, {Skrutskie}, {Weinberg}, \&
  {Ostheimer}}]{maj03}
{Majewski}, S.~R., {Skrutskie}, M.~F., {Weinberg}, M.~D., \& {Ostheimer}, J.~C.
  2003, \apj, 599, 1082

\bibitem[{{Malkin}(2012)}]{malkin12}
{Malkin}, Z. 2012, ArXiv e-prints

\bibitem[{{Martin} {et~al.}(2007){Martin}, {Ibata}, \& {Irwin}}]{martin07}
{Martin}, N.~F., {Ibata}, R.~A., \& {Irwin}, M. 2007, \apjl, 668, L123

\bibitem[{{Newberg} {et~al.}(2002){Newberg}, {Yanny}, {Rockosi}, {Grebel},
  {Rix}, {Brinkmann}, {Csabai}, {Hennessy}, {Hindsley}, {Ibata}, {Ivezi{\'c}},
  {Lamb}, {Nash}, {Odenkirchen}, {Rave}, {Schneider}, {Smith}, {Stolte}, \&
  {York}}]{newb02}
{Newberg}, H.~J., {Yanny}, B., {Rockosi}, C., {et~al.} 2002, \apj, 569, 245

\bibitem[{{Pe{\~n}arrubia} {et~al.}(2010){Pe{\~n}arrubia}, {Belokurov},
  {Evans}, {Mart{\'{\i}}nez-Delgado}, {Gilmore}, {Irwin}, {Niederste-Ostholt},
  \& {Zucker}}]{penarrub10}
{Pe{\~n}arrubia}, J., {Belokurov}, V., {Evans}, N.~W., {et~al.} 2010, \mnras,
  408, L26

\bibitem[{{Pila-D{\'{\i}}ez} {et~al.}(2014){Pila-D{\'{\i}}ez}, {Kuijken}, {de
  Jong}, {Hoekstra}, \& {van der Burg}}]{piladiez14}
{Pila-D{\'{\i}}ez}, B., {Kuijken}, K., {de Jong}, J.~T.~A., {Hoekstra}, H., \&
  {van der Burg}, R.~F.~J. 2014, \aap, 564, A18

\bibitem[{{Rocha-Pinto} {et~al.}(2003){Rocha-Pinto}, {Majewski}, {Skrutskie},
  \& {Crane}}]{rochapinto03}
{Rocha-Pinto}, H.~J., {Majewski}, S.~R., {Skrutskie}, M.~F., \& {Crane}, J.~D.
  2003, \apjl, 594, L115

\bibitem[{{Sand} {et~al.}(2012){Sand}, {Graham}, {Bildfell}, {Zaritsky},
  {Pritchet}, {Hoekstra}, {Just}, {Herbert-Fort}, {Sivanandam}, {Foley}, \&
  {Mahdavi}}]{sand12meneacs}
{Sand}, D.~J., {Graham}, M.~L., {Bildfell}, C., {et~al.} 2012, \apj, 746, 163

\bibitem[{{Schlegel} {et~al.}(1998){Schlegel}, {Finkbeiner}, \&
  {Davis}}]{schlegel98dustmaps}
{Schlegel}, D.~J., {Finkbeiner}, D.~P., \& {Davis}, M. 1998, \apj, 500, 525

\bibitem[{{Sesar} {et~al.}(2010){Sesar}, {Ivezi{\'c}}, {Grammer}, {Morgan},
  {Becker}, {Juri{\'c}}, {De Lee}, {Annis}, {Beers}, {Fan}, {Lupton}, {Gunn},
  {Knapp}, {Jiang}, {Jester}, {Johnston}, \& {Lampeitl}}]{sesar10densProfile}
{Sesar}, B., {Ivezi{\'c}}, {\v Z}., {Grammer}, S.~H., {et~al.} 2010, \apj, 708,
  717

\bibitem[{{Sesar} {et~al.}(2011){Sesar}, {Juri{\'c}}, \&
  {Ivezi{\'c}}}]{sesar11}
{Sesar}, B., {Juri{\'c}}, M., \& {Ivezi{\'c}}, {\v Z}. 2011, \apj, 731, 4

\bibitem[{{Skrutskie} {et~al.}(2006){Skrutskie}, {Cutri}, {Stiening},
  {Weinberg}, {Schneider}, {Carpenter}, {Beichman}, {Capps}, {Chester},
  {Elias}, {Huchra}, {Liebert}, {Lonsdale}, {Monet}, {Price}, {Seitzer},
  {Jarrett}, {Kirkpatrick}, {Gizis}, {Howard}, {Evans}, {Fowler}, {Fullmer},
  {Hurt}, {Light}, {Kopan}, {Marsh}, {McCallon}, {Tam}, {Van Dyk}, \&
  {Wheelock}}]{skrutskie06}
{Skrutskie}, M.~F., {Cutri}, R.~M., {Stiening}, R., {et~al.} 2006, \aj, 131,
  1163

\bibitem[{{Totten} \& {Irwin}(1998)}]{totten98carbonstars}
{Totten}, E.~J. \& {Irwin}, M.~J. 1998, \mnras, 294, 1

\bibitem[{{Watkins} {et~al.}(2009){Watkins}, {Evans}, {Belokurov}, {Smith},
  {Hewett}, {Bramich}, {Gilmore}, {Irwin}, {Vidrih}, {Wyrzykowski}, \&
  {Zucker}}]{watkins09}
{Watkins}, L.~L., {Evans}, N.~W., {Belokurov}, V., {et~al.} 2009, \mnras, 398,
  1757

\bibitem[{{York} {et~al.}(2000){York}, {Adelman}, {Anderson}, {Anderson},
  {Annis}, {Bahcall}, {Bakken}, {Barkhouser}, {Bastian}, {Berman}, {Boroski},
  {Bracker}, {Briegel}, {Briggs}, {Brinkmann}, {Brunner}, {Burles}, {Carey},
  {Carr}, {Castander}, {Chen}, {Colestock}, {Connolly}, {Crocker}, {Csabai},
  {Czarapata}, {Davis}, {Doi}, {Dombeck}, {Eisenstein}, {Ellman}, {Elms},
  {Evans}, {Fan}, {Federwitz}, {Fiscelli}, {Friedman}, {Frieman}, {Fukugita},
  {Gillespie}, {Gunn}, {Gurbani}, {de Haas}, {Haldeman}, {Harris}, {Hayes},
  {Heckman}, {Hennessy}, {Hindsley}, {Holm}, {Holmgren}, {Huang}, {Hull},
  {Husby}, {Ichikawa}, {Ichikawa}, {Ivezi{\'c}}, {Kent}, {Kim}, {Kinney},
  {Klaene}, {Kleinman}, {Kleinman}, {Knapp}, {Korienek}, {Kron}, {Kunszt},
  {Lamb}, {Lee}, {Leger}, {Limmongkol}, {Lindenmeyer}, {Long}, {Loomis},
  {Loveday}, {Lucinio}, {Lupton}, {MacKinnon}, {Mannery}, {Mantsch}, {Margon},
  {McGehee}, {McKay}, {Meiksin}, {Merelli}, {Monet}, {Munn}, {Narayanan},
  {Nash}, {Neilsen}, {Neswold}, {Newberg}, {Nichol}, {Nicinski}, {Nonino},
  {Okada}, {Okamura}, {Ostriker}, {Owen}, {Pauls}, {Peoples}, {Peterson},
  {Petravick}, {Pier}, {Pope}, {Pordes}, {Prosapio}, {Rechenmacher}, {Quinn},
  {Richards}, {Richmond}, {Rivetta}, {Rockosi}, {Ruthmansdorfer}, {Sandford},
  {Schlegel}, {Schneider}, {Sekiguchi}, {Sergey}, {Shimasaku}, {Siegmund},
  {Smee}, {Smith}, {Snedden}, {Stone}, {Stoughton}, {Strauss}, {Stubbs},
  {SubbaRao}, {Szalay}, {Szapudi}, {Szokoly}, {Thakar}, {Tremonti}, {Tucker},
  {Uomoto}, {Vanden Berk}, {Vogeley}, {Waddell}, {Wang}, {Watanabe},
  {Weinberg}, {Yanny}, {Yasuda}, \& {SDSS Collaboration}}]{york00}
{York}, D.~G., {Adelman}, J., {Anderson}, Jr., J.~E., {et~al.} 2000, \aj, 120,
  1579

\end{thebibliography}

%\begin{thebibliography}{}

  %\bibitem[1966]{baker} Baker, N. 1966,
  %    in Stellar Evolution,
  %    ed.\ R. F. Stein,\& A. G. W. Cameron
  %    (Plenum, New York) 333

   %\bibitem[1988]{balluch} Balluch, M. 1988,
   %   A\&A, 200, 58

   %\bibitem[1980]{cox} Cox, J. P. 1980,
   %   Theory of Stellar Pulsation
   %   (Princeton University Press, Princeton) 165

   %\bibitem[1969]{cox69} Cox, A. N.,\& Stewart, J. N. 1969,
   %   Academia Nauk, Scientific Information 15, 1

   %\bibitem[1980]{mizuno} Mizuno H. 1980,
   %   Prog. Theor. Phys., 64, 544
   
   %\bibitem[1987]{tscharnuter} Tscharnuter W. M. 1987,
   %   A\&A, 188, 55
  
   %\bibitem[1992]{terlevich} Terlevich, R. 1992, in ASP Conf. Ser. 31, 
   %   Relationships between Active Galactic Nuclei and Starburst Galaxies, 
   %   ed. A. V. Filippenko, 13

   %\bibitem[1980a]{yorke80a} Yorke, H. W. 1980a,
   %   A\&A, 86, 286

   %\bibitem[1997]{zheng} Zheng, W., Davidsen, A. F., Tytler, D. \& Kriss, G. A.
   %   1997, preprint
%\end{thebibliography}

%%%%%%%%%%%%%%%%%%%%%%%%%%%%%%%%%%%%%%%%%%%%%%%%%%%%%%%%%%%%%%%%%%%%%%%%%%%%%%%%%%%%%%%%%%%%
%  _________________________________________________________________________________________
%  _________________________________________________________________________________________
%
%%%%%%%%%%%%%%%%%%%%%%%%%%%%%%%%%%%%%%%%%%%%%%%%%%%%%%%%%%%%%%%%%%%%%%%%%%%%%%%%%%%%%%%%%%%%

\end{document}